%% file: half-KFN.tex
\documentclass[journal]{IEEEtran}
\usepackage{amsfonts,amsmath}
\usepackage{textcomp}
\usepackage{stfloats}
\usepackage{url}
\usepackage{verbatim}
\usepackage{graphicx}
\RequirePackage[colorlinks,citecolor=blue,urlcolor=blue]{hyperref}
\usepackage{indentfirst}
\usepackage{psfrag,epsf}
\usepackage{enumerate}
\RequirePackage{amsthm}
\usepackage{amssymb}
\usepackage{booktabs}
\usepackage{bbm}
\usepackage{color}
\usepackage{siunitx}
\usepackage{subfigure}
\usepackage{makecell}%oblique line in the table
\usepackage{arydshln}
\usepackage{diagbox}
\usepackage{multirow}
\usepackage{array}
\usepackage{float}
\usepackage{cleveref}
\usepackage{algorithm}
\usepackage{algorithmic}
\usepackage{tikz}
\usepackage{threeparttable}
\usetikzlibrary{decorations.pathreplacing,calligraphy}
\usetikzlibrary{arrows,positioning,fit,shapes,calc}
\usepackage{dcolumn}%
\newtheorem{proposition}{Proposition}

\newtheorem{thm}{Theorem}[section]
\newcommand{\bee} {\begin{equation} }
\newcommand{\ene}{\end{equation}}

 \numberwithin{equation}{section}
 \numberwithin{lem}{section}

\DeclareGraphicsExtensions{.pdf,.png,.jpg}

\makeatletter%
\@addtoreset{equation}{section}

\newcommand\figcaption{\def\@captype{figure}\caption}
\newcommand\tabcaption{\def\@captype{table}\caption}
\makeatother%

\begin{document}
% \bibliographystyle{IEEEtran}
% \def\spacingset#1{\renewcommand{\baselinestretch}%
% {#1}\small\normalsize} %\spacingset{1}
%%%%%%%%%%%%%%%%%%%%%%%%%%%%%%%%%%%%%%%%%%%%%%%%%%%%%%%%%%%%%%%%%%%%%%%%%%%%%%
% \if0\blind
\title{Half-KFN: An Enhanced  Detection Method for Subtle Covariate Drift} 
\author{ Bingbing Wang, Dong Xu, Yu Tang 
\thanks{This work was supported in part by the National Natural Science Foundation of China under Grant 12071329 and Grant 12471246. \textit{(Corresponding authors: Dong Xu.)}}
\thanks{Bingbing Wang and Dong Xu are with the School of Mathematical Sciences, Soochow University, Suzhou 215031, China (e-mail: bbwangstat1@stu.suda.edu.cn;  xd0309@suda.edu.cn).}
\thanks{Yu Tang is with the School of Future Science and Engineering, Soochow University, Suzhou 215006, China (e-mail: ytang@suda.edu.cn).}
}

\maketitle
% \let\thefootnote\relax\footnotetext{* Correspondence: xd0309@suda.edu.cn}
% \let\thefootnote\relax\footnotetext{\textsuperscript{\dag} This work is partically supported by NNSF of China with grant 12071329 and 12471246.}
% }
% \fi
% \if1\blind
% {
%   \bigskip
%   \bigskip
%   \bigskip
%   \begin{center}
%     {\LARGE\bf Title}
% \end{center}
%   \medskip
% } \fi
% \bigskip

\begin{abstract}
Detecting covariate drift is a common task of significant practical value in supervised learning. Once covariate drift occurs, the models may no longer be applicable, hence numerous studies have been devoted to the advancement of detection methods. However, current research methods are not particularly effective in handling subtle covariate drift when dealing with small proportions of drift samples. In this paper, inspired by the $k$-nearest neighbor (KNN) approach, a novel method called Half $k$-farthest neighbor (Half-KFN) is proposed in response to specific scenarios. Compared to traditional ones, Half-KFN exhibits higher power due to the inherent capability of the farthest neighbors which could better characterize the nature of drift. Furthermore, with larger sample sizes, the employment of the bootstrap for hypothesis testing is recommended. It is leveraged to calculate $p$-values dramatically faster than permutation tests, with speed undergoing an exponential growth as sample size increases. Numerical experiments on simulated and real data are conducted to evaluate our proposed method, and the results demonstrate that it consistently displays superior sensitivity and rapidity in covariate drift detection across various cases. 
\end{abstract}

% {\it \textbf{Keywords}: Bootstrap hypothesis test; Covariate drift detection; $k$-nearest neighbor; Permutation test.}     %3 to 6 keywords, that do not appear in the title
% % \vfill

% \newpage
% \spacingset{1.5} % DON'T change the spacing!

\section{Introduction}
\label{section:1}
\IEEEPARstart{N}{owadays}, supervised learning models play a critical role in the field of machine learning, with numerous research efforts dedicated to developing a variety of application tools to drive their progress \cite{nasteski2017overview}, \cite{van2020survey}, \cite{zhou2018brief}. Typically, the assumption of supervised learning is that the training data (following distribution $p$) and the test data (following distribution $q$) are independently and identically distributed, meaning that they have the same stationary distribution, i.e., $p(X, Y) = q(X, Y)$. When this assumption does not hold, we say that dataset drift has occurred \cite{moreno2012unifying}, \cite{polo2023unified}. Dataset drift can be classified based on whether the alteration is in covariate distribution, label distribution, or the relationship between covariates and labels \cite{cobb2022context}, \cite{liu2020concept}, \cite{wang2020multiscale}. Among them, covariate drift caused by changes in covariate distribution is one of the common and significant issues in dataset drift \cite{shimodaira2000improving}. It refers to the situation where only when the marginal distribution of the covariate $p(X)$ deviates from $q(X)$ (i.e. $p(X) \neq q(X)$, while $p(Y|X)= q(Y|X)$). The occurrence of covariate drift may compromise the generalization ability of a trained model, increasing the risk of erroneous predictions. Therefore, covariate drift remains a significant challenge in model deployment, and a substantial amount of research has been devoted to proposing detection methods aimed at helping to promptly identify the incidence of covariate drift \cite{awais2020revisiting}, \cite{wilson2020survey}.

Goodfellow et al. \cite{Goodfellow2014ExplainingAH} developed a method to generate adversarial examples by adding a specially designed small noise to the pixel values of original images, which could result in significant changes in prediction outcomes. Moreover, even if the proportion of drifting samples is subtle, they can become a fatal hidden danger that cannot be ignored for scenarios that rely heavily on models for precise predictions or critical decision-making. Such subtle covariate drift is sufficient to shake the robustness of models, seriously jeopardizing prediction accuracy and decision-making effectiveness. Nevertheless, the detection of subtle covariate drift has not been adequately emphasized. How to identify subtle covariate drift more quickly and accurately poses a challenging problem, especially in the case of large sample sizes. Thus, the motivation behind this study is based on this issue. 
% Dataset drift primarily involves the following three scenarios \cite{moreno2012unifying} \cite{quinonero2022dataset}: (1) covariate drift: only when the marginal distribution of the covariate $p(X)$ deviates from $q(X)$ (i.e. $p(X) \neq q(X)$, while $p(Y|X)= q(Y|X)$), (2) label drift \cite{garg2020unified} \cite{lipton2018detecting} \cite{2022Domain}: only when the marginal distribution of the label changes (i.e. $p(Y) \neq q(Y)$, while $p(X|Y)= q(X|Y)$), (3) concept drift \cite{lu2018learning} \cite{zhang2013domain}: only when the conditional distribution $p(X|Y)$ differs from $q(X|Y)$  (i.e. $p(X|Y) \neq q(X|Y)$, while $p(Y)=q(Y)$). One important challenge in current research is the detection of concept drift within data streams \cite{suarez2023survey} \cite{wang2015concept} \cite{yu2019concept}.

The contributions of this paper are as follows: 
\begin{enumerate}[(1)]
    \item A novel, ingenious, and efficient method called Half-KFN is proposed, which demonstrates higher power compared to the traditional methods for covariate drift detection.
    \item In scenarios with large sample sizes, the faster establishment of critical values compared to permutation tests is enabled by the use of bootstrap hypothesis testing, thereby facilitating the quicker detection of covariate drift. 
    \item Multiple comparative experiments are conducted covering different sample sizes, drift proportions, and various types of covariate drift applied to the test samples. Compared to conventional detection methods, our proposed Half-KFN combined with bootstrap hypothesis test demonstrates multifaceted superiority.
\end{enumerate}

The remaining sections of this paper are organized as follows. Firstly, in Section \ref{section:Related work}, we provide an overview of the latest research relevant to our study. Next, in Section \ref{section:Methodology}, we introduce a covariate drift detection framework based on the proposed Half-KFN statistic, employing bootstrap for expedited hypothesis testing in scenarios with large sample sizes. To demonstrate the effectiveness of our method in various scenarios, we apply this covariate drift framework to both artificial and real-world datasets in Section \ref{section:Experimental evaluation}. Finally, we broadly discuss our research findings and contributions in section \ref{section:Conclusion and Discussion}. The theoretical proof in the paper is placed in the appendix.

% \subsection{Contributions}
% (1) Our work primarily focuses on the data drift detection component, where we improve upon the $k$-nearest neighbors (KNN) test by employing our Half-KFN method. For slightly proportioned drifting samples, their predictive performance deteriorates through classifier $f$, possibly drifting towards other distributions. Cases of larger distances are more likely to depict this degradation in predictive accuracy, hence the $k$-farthest neighbors are more applicable to such specific scenarios.

% (2) We introduce two detection methods that can be selected based on the sample size: the permutation test approach excels in accurately detecting scenarios involving minor drift and small sample sizes, while maintaining power control under the null hypothesis. Conversely, the bootstrap hypothesis test demonstrates proficiency in accurately identifying cases involving more subtle drift and larger sample sizes, while also ensuring power control under the null hypothesis, thus enhancing the reliability and sensitivity of detection.

\section{Related work}
\label{section:Related work}
\subsection{Covariate drift detection}
\label{covariate drift detection}
Covariate drift detection methods are typically categorized based on different sample formats, which can be classified into streaming/online mode \cite{raza2016adaptive}, \cite{raza2014adaptive} and two-sample tests in a batch/offline mode \cite{rabanser2019failing}, \cite{sisniega2024efficient}. To address the first mode, Xu et al. \cite{xu2021concept} propose using multiple signals to capture a wide range of characteristics of the data. In addition, Polo et al. \cite{maia2023effective} demonstrated the importance of dimensionality reduction or feature selection for model generalization in the context of covariate shift adaptation. Certain methods assist in identifying issues such as model performance degradation caused by changes in data distribution, promptly triggering alerts \cite{kore2024empirical}, \cite{osti_10334960}. Some methods specifically focus on the covariate drift correction to improve the robustness and generalization ability of the model \cite{chan2020unlabelled}, \cite{park2020calibrated}, \cite{DBLPjournalscorrabs211004065}. To address the second mode, Rabanser et al. \cite{rabanser2019failing} conducted a comparative study on various dimensionality reduction methods using traditional two-sample tests in the context of covariance shift detection. By enhancing the detection framework, Castle e al. \cite{castle2021detecting} represented the data as model explanations in the form of gradient times input, providing additional information about the data domain for two-sample tests, which helps improve the sensitivity of drift detection.

\subsection{A dimension reduction method}
\label{A dimension reduction method}
In the current context of supervised learning, input data is commonly characterized by high dimensionality, which presents challenges. To address the detection of covariate drift in high-dimensional input data, Lipton et al. \cite{lipton2018detecting} proposed a method called black box drift detection (BBSD) which is related to detect covariate drift \cite{rabanser2019failing}, concept drift \cite{wijaya2021failing}, label drift \cite{mukherjee2022detecting} and more
general forms of nonstationarity. Furthermore, Rabanser et al. \cite{rabanser2019failing} propose using the softmax outputs (BBSDs) of a label classifier, trained on source data, as a dimension-reduced representation. They also found that compared to various dimensionality reduction methods such as principal components analysis (PCA), sparse random projection (SRP), untrained autoencoders (UAE), and trained autoencoders (TAE), using BBSDs can achieve the best performance. It simply requires the utilization of an arbitrary soft-classifier $f$ (required that the confusion matrix constructed from the classification results on the training data is invertible) to detect drift in the distributions of each class after dimensionality reduction of the covariates, i.e., $p(f(X)) \neq q(f(X))$.

\subsection{Two-sample test}
\label{Two-sample test}
Once the dimensionality reduction method has been determined, it is natural to proceed with the detection method. The multivariate two-sample test has been extensively investigated in studies, with notable contributions from research papers that have explored methods such as the Maximum Mean Discrepancy (MMD) test \cite{biggs2024mmd}, \cite{gretton2012kernel}, \cite{schrab2023mmd}, Friedman-Rafsky (FR) test \cite{friedman1979multivariate}, \cite{hsiao2016mapping}, and energy test \cite{aslan2005statistical}, \cite{zech2018scaling}. Moreover, KNN test proposed by Schilling \cite{schilling1986multivariate} and Henze \cite{1988A} is a famous non-parametric method for multivariate two-sample test. Djolonga and Krause \cite{djolonga2017learning} devised differentiable adaptations of the KNN test, thereby establishing the asymptotic normality under specific conditions. By smoothing classical KNN statistic, the computation of smooth $p$-values becomes feasible under permutation distribution. Although KNN test has the asymptotic normality under large samples \cite{1988A}, \cite{mondal2015high}, \cite{schilling1986multivariate}, \cite{schilling1986mutual}, the detection effect is not satisfactory in special scenarios with small proportion drifting samples. For this reason, current methods for detecting covariate drift still have limitations, which highlight the crucial necessity for research aimed at accurately identifying subtle covariate drift.

\subsection{Resampling hypothesis test}
\label{Resampling Hypothesis Test}
In the statistical inference task of a two-sample test, resampling methods are flexible and powerful tools, which involve making repeated random samples from a dataset for statistical inference \cite{chihara2022mathematical}, \cite{lock2020statistics}. Resampling hypothesis test, which includes permutation test and bootstrap hypothesis test, is utilized to estimate the distribution of statistics, calculate confidence intervals, assist in assessing the robustness of parameter estimates, and more. These techniques demonstrate significant potential and impact in practical applications.

When the population distribution is unclear, one method for statistical inference is permutation test \cite{diciccio2020evaluating}, \cite{good2013permutation}, \cite{polo2023unified}. One of the most significant advantages of using permutation methods for inference is their robustness \cite{lafleur2009introduction}. Permutation tests have relatively weak assumptions about the statistical data, making them more robust than corresponding parametric methods when dealing with outliers or extreme distributions. Permutation tests are particularly effective in tests with small sample sizes \cite{frossard2021permutation}, \cite{yu2024robust}. However, a potential issue arises as the sample size increases, leading to longer testing times \cite{shekhar2023permutation}. Another detection method involves obtaining critical values through bootstrap under the null hypothesis \cite{hesterberg2015teachers}, \cite{yolcu2021probabilistic}, where the Central Limit Theorem ensures the asymptotic normality of the mean of the test statistic in large sample scenarios. If the sample size is sufficiently large, even if the distribution of the original data is non-normal, parametric tests can still be employed \cite{chernick2014introduction}, \cite{dwivedi2017analysis}.
% \section{Problem setup}
% \label{section:Problem setup}
% Problem setup

\section{Methodology}
\label{section:Methodology}
Our approach highlights two notable innovations. Firstly, after dimensionality reduction through BBSDs, we craft the Half-KFN statistic using Euclidean distance. Based on the characteristics of farthest neighbors, this statistic can more precisely represent the magnitude of differences arising from small proportions of covariate drifting samples. Secondly, we observe significant computation time and resource consumption issues when performing permutation testing to calculate $p$-values in statistical inference involving larger sample sizes. So, we propose employing bootstrap to construct a new statistic defined as the mean of Half-KFN, and we calculate its expectation and asymptotic variance for statistical inference.

Experimental results demonstrate that our method exhibits advantages in various aspects. It is not only capable of conducting statistical inference more rapidly and accurately, but also consumes less computational resources. Furthermore, it effectively controls the probability of Type-I error.

% Our approach is both simple and efficient. Given any BBSDs classifier, we perform dimensionality reduction on the covariates and subsequently use Euclidean distance to construct and define the Half-KFN statistic, which is employed to characterize the magnitude of differences caused by covariate drift. 

% The focus is on the fact that, in large sample scenarios, permutation tests for $p$-value computation are time-consuming and resource-intensive. So, we propose a more efficient and faster method that utilizes bootstrap to construct the mean of the Half-KFN statistic. According to the central limit theorem, when the sample size is sufficiently large, the mean of the statistic will approximately follow a normal distribution. Using this method, we only need to calculate the expectation and variance of the mean of the statistic to conduct statistical inference. Furthermore, this method can effectively control the probability of Type-I error. The results indicate that, compared to permutation tests, our method offers the advantage of performing statistical inference quickly and accurately with fewer computational resources. Additionally, compared to other traditional detection methods, our approach demonstrates superior detection efficiency.

\tikzset{
every transaction/.style = {fill=white!100},
transaction/.style = { draw, minimum size=6mm, every transaction},
every actor role/.style = {},
actor role/.style = {rectangle, draw=black!80, ultra thick,
    minimum size = 6mm, every actor role},
composite actor role/.style = {fill=pink!80, actor role},
composite actor role2/.style = {fill=gray!80, actor role,line width = 1pt},
composite actor role3/.style = {fill=yellow!80, actor role,line width = 0.1pt},
elementary actor role/.style = {fill=white!100, actor role},
initiator/.style = {-},
executor/.style = {<-, >=},
system/.style = {rectangle, fill=white!100, ultra thick, draw=red!80,
            minimum height=60mm, minimum width=4.5cm,outer sep=0pt,
            dash pattern = on 2pt off 2pt on 2pt off 2pt,line width = 0.5pt}}

\begin{figure}[!t]
\centering
\scalebox{0.57}{
\pgfdeclarelayer{background}
\pgfdeclarelayer{foreground}
\pgfsetlayers{background,main,foreground}
\begin{tikzpicture}[node distance=1cm, on grid]
    \begin{pgfonlayer}{background}
        \node [system] (system) at (2.5,3){};
        \node [above] at (system.north) {Drift Detection};
    \end{pgfonlayer}
    \node [composite actor role2] (BBSDs) [minimum height=48mm] at ( -4,3) {BBSDs};
    \node [above] at (BBSDs.north) {Dimension Reduction};
    \path (BBSDs)++(-2,1) node [transaction] (Xsource) {${X}_{source}$} ;
    \node [above] at (Xsource.north) {Input};
    \draw[thick,gray,->] (Xsource.east) -- ($(BBSDs.south west)!.706!(BBSDs.north west)$);
    \path (BBSDs)++(-2,-1) node [transaction] (Xtarget) {${X}_{target}$};
    \draw[thick,gray,->] (Xtarget.east) -- ($(BBSDs.south west)!.294!(BBSDs.north west)$);
    % \node [transaction] (ysource) at(-2,4)
    %     {$\hat{y}_{source}$} edge [initiator] (BBSDs.south east |- ysource);
    \node [transaction] (ysource) at(-1.5,4) {$\hat{Y}_{source}$};
    \node [above] at (ysource.north) {Feature Extraction};
    \draw[thick,gray,->] ($(BBSDs.south east)!.706!(BBSDs.north east)$) -- (ysource.west);
    \draw[thick,gray,->] (ysource.east) -- ($(system.south west)!.6667!(system.north west)$);
    \node [composite actor role] (Half-KFN) at ( $(system.south)!.8!(system.north)$)
        {Half-KFN} ;
    \draw[thick,red,->] (3.1,4.5) -- (3.1,3.8);
    \node [above] at (Half-KFN.north) {$\color{red}{\star}$Statistic};
    \node [transaction] (output) at (7,3) {\text{Detection Result}}  ;
    % \node [composite actor role3] (PT) at (7,4) {\text{\scriptsize Permutation Test}} ;
    % \node [composite actor role3] (AND) at (7,2) {\text{\scriptsize Asymptomatic Null Distribution}};
    % \draw[thick,gray,->] ($(system.south east)!.66667!(system.north east)$) -- (PT.west);
    % \draw[thick,gray,->] ($(system.south east)!.3333!(system.north east)$) -- (AND.west);
    % \draw[thick,gray,->] ($(PT.south)$) -- (output);
    % \draw[thick,gray,->] ($(AND.north)$) -- (output);
    \draw[thick,gray,->] ($(system.east)$) -- (output);
    \node [above] at (output.north) {\text{Output}};
    \node [transaction] (ytarget) at (-1.5,2) {$\hat{Y}_{target}$} ;
    \draw[thick,gray,->] ($(BBSDs.south east)!.294!(BBSDs.north east)$) -- (ytarget.west);
    \draw[thick,gray,->] (ytarget.east) -- ($(system.south west)!.3333!(system.north west)$);
    \node [composite actor role3] at (2.5,1.1) {\text{\scriptsize Permutation Test}};
    \node [composite actor role3] at (2.5,0.4) {\text{\scriptsize Bootstrap Hypothesis Test}};
        \draw[red, xshift=1cm, yshift=1.5cm, scale=0.35](6,0.000)--(6,6.008);
        \draw[orange, xshift=1cm, yshift=1.5cm, scale=0.35] (-0.022,0.000)--(-0.022,0.000);
        \draw[orange, xshift=1cm, yshift=1.5cm, scale=0.35] (0.000,0.000)--(0.022,0.000)--(0.067,0.000)--(0.114,0.000)%
          --(0.159,0.000)--(0.204,0.000)--(0.249,0.000)--(0.294,0.000)--(0.340,0.000)%
          --(0.385,0.000)--(0.430,0.000)--(0.475,0.000)--(0.520,0.000)--(0.566,0.000)%
          --(0.611,0.000)--(0.656,0.000)--(0.701,0.000)--(0.746,0.000)--(0.792,0.000)%
          --(0.792,0.227)--(0.837,0.227)--(0.882,0.227)--(0.927,0.227)--(0.972,0.227)%
          --(1.018,0.227)--(1.063,0.227)--(1.108,0.227)--(1.153,0.227)--(1.198,0.227)%
          --(1.244,0.227)--(1.289,0.227)--(1.334,0.227)--(1.334,1.134)--(1.379,1.134)%
          --(1.424,1.134)--(1.470,1.134)--(1.515,1.134)--(1.560,1.134)--(1.605,1.134)%
          --(1.650,1.134)--(1.696,1.134)--(1.741,1.134)--(1.786,1.134)--(1.831,1.134)%
          --(1.831,2.836)--(1.876,2.836)--(1.923,2.836)--(1.968,2.836)--(2.013,2.836)%
          --(2.058,2.836)--(2.103,2.836)--(2.149,2.836)--(2.194,2.836)--(2.239,2.836)%
          --(2.284,2.836)--(2.329,2.836)--(2.375,2.836)--(2.375,4.727)--(2.420,4.727)%
          --(2.465,4.727)--(2.510,4.727)--(2.555,4.727)--(2.601,4.727)--(2.646,4.727)%
          --(2.691,4.727)--(2.736,4.727)--(2.781,4.727)--(2.827,4.727)--(2.872,4.727)%
          --(2.917,4.727)--(2.917,5.909)--(2.962,5.909)--(3.007,5.909)--(3.053,5.909)%
          --(3.098,5.909)--(3.143,5.909)--(3.188,5.909)--(3.233,5.909)--(3.279,5.909)%
          --(3.324,5.909)--(3.369,5.909)--(3.414,5.909)--(3.459,5.909)--(3.505,5.909)%
          --(3.550,5.909)--(3.595,5.909)--(3.640,5.909)--(3.685,5.909)--(3.732,5.909)%
          --(3.777,5.909)--(3.822,5.909)--(3.867,5.909)--(3.912,5.909)--(3.958,5.909)%
          --(3.958,4.924)--(4.003,4.924)--(4.048,4.924)--(4.093,4.924)--(4.138,4.924)%
          --(4.184,4.924)--(4.229,4.924)--(4.274,4.924)--(4.319,4.924)--(4.364,4.924)%
          --(4.410,4.924)--(4.455,4.924)--(4.500,4.924)--(4.500,3.517)--(4.545,3.517)%
          --(4.590,3.517)--(4.636,3.517)--(4.681,3.517)--(4.726,3.517)--(4.771,3.517)%
          --(4.816,3.517)--(4.862,3.517)--(4.907,3.517)--(4.952,3.517)--(4.997,3.517)%
          --(5.042,3.517)--(5.042,2.199)--(5.088,2.199)--(5.133,2.199)--(5.178,2.199)%
          --(5.223,2.199)--(5.268,2.199)--(5.314,2.199)--(5.359,2.199)--(5.404,2.199)%
          --(5.449,2.199)--(5.494,2.199)--(5.541,2.199)--(5.541,1.221)--(5.586,1.221)%
          --(5.631,1.221)--(5.676,1.221)--(5.721,1.221)--(5.767,1.221)--(5.812,1.221)%
          --(5.857,1.221)--(5.902,1.221)--(5.947,1.221)--(5.993,1.221)--(6.038,1.221)%
          --(6.083,1.221)--(6.083,0.611)--(6.128,0.611)--(6.173,0.611)--(6.219,0.611)%
          --(6.264,0.611)--(6.309,0.611)--(6.354,0.611)--(6.399,0.611)--(6.445,0.611)%
          --(6.490,0.611)--(6.535,0.611)--(6.580,0.611)--(6.625,0.611)--(6.625,0.278)%
          --(6.671,0.278)--(6.716,0.278)--(6.761,0.278)--(6.806,0.278)--(6.851,0.278)%
          --(6.897,0.278)--(6.942,0.278)--(6.987,0.278)--(7.032,0.278)--(7.077,0.278)%
          --(7.123,0.278)--(7.168,0.278)--(7.168,0.115)--(7.213,0.115)--(7.258,0.115)%
          --(7.303,0.115)--(7.350,0.115)--(7.395,0.115)--(7.440,0.115)--(7.485,0.115)%
          --(7.530,0.115)--(7.576,0.115)--(7.621,0.115)--(7.666,0.115)--(7.666,0.045)%
          --(7.711,0.045)--(7.756,0.045)--(7.802,0.045)--(7.847,0.045)--(7.892,0.045)%
          --(7.937,0.045)--(7.982,0.045)--(8.028,0.045)--(8.073,0.045)--(8.118,0.045)%
          --(8.163,0.045)--(8.208,0.045)--(8.208,0.016)--(8.254,0.016)--(8.299,0.016)%
          --(8.344,0.016)--(8.389,0.016)--(8.434,0.016)--(8.480,0.016)--(8.525,0.016)%
          --(8.570,0.016)--(8.615,0.016)--(8.660,0.016)--(8.706,0.016)--(8.751,0.016)%
          --(8.751,0.005)--(8.796,0.005)--(8.841,0.005)--(8.886,0.005)--(8.932,0.005)%
          --(8.977,0.005)--(9.000,0.005);
        \draw[green, xshift=1cm, yshift=1.5cm, scale=0.35] (0.000,0.045)--(0.045,0.051)--(0.090,0.056)--(0.136,0.064)%
          --(0.181,0.071)--(0.226,0.080)--(0.271,0.090)--(0.316,0.099)--(0.362,0.111)%
          --(0.407,0.124)--(0.452,0.138)--(0.497,0.153)--(0.542,0.169)--(0.588,0.187)%
          --(0.633,0.207)--(0.678,0.228)--(0.723,0.251)--(0.768,0.277)--(0.814,0.304)%
          --(0.859,0.333)--(0.904,0.366)--(0.949,0.400)--(0.994,0.437)--(1.041,0.476)%
          --(1.086,0.518)--(1.131,0.564)--(1.176,0.612)--(1.221,0.664)--(1.267,0.719)%
          --(1.312,0.777)--(1.357,0.839)--(1.402,0.905)--(1.447,0.973)--(1.493,1.046)%
          --(1.538,1.123)--(1.583,1.203)--(1.628,1.288)--(1.673,1.376)--(1.719,1.468)%
          --(1.764,1.564)--(1.809,1.664)--(1.854,1.768)--(1.899,1.875)--(1.945,1.986)%
          --(1.990,2.102)--(2.035,2.219)--(2.080,2.340)--(2.125,2.465)--(2.171,2.592)%
          --(2.216,2.722)--(2.261,2.854)--(2.306,2.988)--(2.351,3.123)--(2.397,3.261)%
          --(2.442,3.398)--(2.487,3.538)--(2.532,3.677)--(2.577,3.816)--(2.623,3.955)%
          --(2.668,4.092)--(2.713,4.228)--(2.758,4.363)--(2.803,4.495)--(2.850,4.625)%
          --(2.895,4.751)--(2.940,4.873)--(2.985,4.991)--(3.030,5.105)--(3.076,5.214)%
          --(3.121,5.317)--(3.166,5.415)--(3.211,5.506)--(3.256,5.590)--(3.302,5.668)%
          --(3.347,5.738)--(3.392,5.801)--(3.437,5.856)--(3.482,5.902)--(3.528,5.940)%
          --(3.573,5.970)--(3.618,5.992)--(3.663,6.004)--(3.708,6.008)--(3.754,6.003)%
          --(3.799,5.990)--(3.844,5.968)--(3.889,5.936)--(3.934,5.897)--(3.980,5.849)%
          --(4.025,5.794)--(4.070,5.730)--(4.115,5.659)--(4.160,5.581)--(4.206,5.495)%
          --(4.251,5.403)--(4.296,5.305)--(4.341,5.201)--(4.386,5.092)--(4.433,4.978)%
          --(4.478,4.858)--(4.522,4.736)--(4.567,4.609)--(4.614,4.480)--(4.659,4.347)%
          --(4.704,4.213)--(4.749,4.076)--(4.794,3.938)--(4.840,3.800)--(4.885,3.660)%
          --(4.930,3.521)--(4.975,3.383)--(5.020,3.244)--(5.066,3.107)--(5.111,2.971)%
          --(5.156,2.838)--(5.201,2.706)--(5.246,2.577)--(5.292,2.450)--(5.337,2.326)%
          --(5.382,2.205)--(5.427,2.088)--(5.472,1.973)--(5.518,1.863)--(5.563,1.756)%
          --(5.608,1.652)--(5.653,1.553)--(5.698,1.458)--(5.744,1.365)--(5.789,1.278)%
          --(5.834,1.194)--(5.879,1.113)--(5.924,1.037)--(5.970,0.965)--(6.015,0.896)%
          --(6.060,0.832)--(6.105,0.770)--(6.150,0.712)--(6.196,0.657)--(6.241,0.606)%
          --(6.286,0.558)--(6.331,0.513)--(6.376,0.471)--(6.423,0.432)--(6.468,0.395)%
          --(6.513,0.361)--(6.558,0.330)--(6.603,0.301)--(6.649,0.274)--(6.694,0.249)%
          --(6.739,0.226)--(6.784,0.205)--(6.829,0.185)--(6.875,0.167)--(6.920,0.151)%
          --(6.965,0.136)--(7.010,0.122)--(7.055,0.109)--(7.101,0.099)--(7.146,0.088)%
          --(7.191,0.079)--(7.236,0.070)--(7.281,0.063)--(7.327,0.056)--(7.372,0.050)%
          --(7.417,0.044)--(7.462,0.039)--(7.507,0.035)--(7.553,0.031)--(7.598,0.027)%
          --(7.643,0.024)--(7.688,0.021)--(7.733,0.018)--(7.779,0.016)--(7.824,0.014)%
          --(7.869,0.012)--(7.914,0.011)--(7.959,0.009)--(8.005,0.008)--(8.050,0.007)%
          --(8.095,0.006)--(8.140,0.005)--(8.185,0.005)--(8.232,0.004)--(8.277,0.003)%
          --(8.322,0.003)--(8.367,0.002)--(8.412,0.002)--(8.458,0.002)--(8.503,0.002)%
          --(8.548,0.002)--(8.593,0.001)--(8.638,0.001)--(8.684,0.001)--(8.729,0.001)%
          --(8.774,0.001)--(8.819,0.001)--(8.864,0.001)--(8.910,0.000)--(8.955,0.000)--(9.000,0.000);
\end{tikzpicture}
}
\\[2em]
    \caption{\small Data drift detection architecture with Half-KFN. First, data are respectively sampled from source and target distributions. Subsequently, the BBSDs method is applied for performing dimensionality reduction. Then, our proposed statistical measure, Half-KFN, is utilized to quantify the magnitude of drift differences. Finally, permutation test and bootstrap hypothesis test is employed to obtain the corresponding $p$-value.}
    \label{fig:1}
\end{figure}
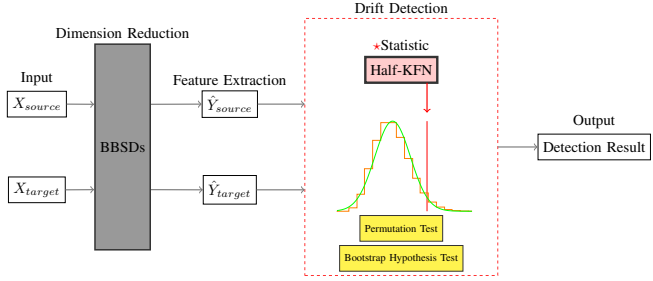

\subsection{Problem setup}\label{Problem setup}
Let $\{X_1,...,X_{n_1}\}$ and $\{X'_{1},...,X'_{n_2}\}$ be two datasets of covariates from the source distribution $p$ and the target distribution $q$ in $\mathbb{R}^{C}$, respectively. 
We make the assumption of independent and identically distributed (i.i.d.) observations within each dataset, and further assume that the datasets are mutually independent.

The problem of covariate drift detection here is to test the hypothesis $H_0:p(X)=q(X)$ against the completely general alternative $H_1:p(X) \neq q(X)$. In the context of subtle covariate shift detection, our primary concerns encompass two core questions: (1) How to sensitively capture the drifted samples? (2) How to perform the detection faster with larger sample sizes?

\subsection{Dimension Reduction}\label{Dimension Reduction}
High-dimensional covariate input data usually requires efficient dimensionality reduction techniques to simplify data representation and extract key information. It is worth noting that without making assumptions about the data, these mappings, which reduce the dimensionality from $C$ to $L$ (with $C > L$), are typically surjective, meaning each output element has at least one corresponding input, and multiple inputs can map to the same output. Although it is possible to construct special cases where the input distribution changes while the low-dimensional latent representation distribution remains unchanged, such situations are likely very rare in practical non-adversarial environments. Hence, dimensionality reduction techniques are practical and can effectively simplify high-dimensional data and extract key information.

Lipton et al. \cite{lipton2018detecting} proposed a method called Black Box drift Detection (BBSDs), demonstrating that if one has an arbitrary pre-trained label classifier $f$, and its confusion matrix on the source distribution data is invertible, detecting whether covariate drift has occurred is equivalent to detecting whether $p(f(X))$ equals $q(f(X))$. It states that for any fixed measurable function $f: \mathbb{R}^{C} \rightarrow \mathbb{R}^{L}$ and for every measurable $\mathcal{S}\subset\mathbb{R}^{L}$, 
\begin{equation}
    p(f(X))=q(f(X))\Rightarrow p(X \in f^{-1}(\mathcal{S}))=q(X \in f^{-1}(\mathcal{S})).
\end{equation}

Rabanser et al. \cite{rabanser2019failing} explored a series of dimensionality reduction techniques and conducted comparative experiments under various simulated covariate drift scenarios. The experiments showed that BBSDs performed best in dimensionality reduction across various drift simulations. Therefore, in this paper, we will use the BBSDs method for dimensionality reduction.

Assume there is a label classifier $f$ whose confusion matrix is invertible and is a mapping from $\mathbb{R}^{C}$ to $\mathbb{R}^{L}$. Let the BBSDs dimensionality reduction result be $\hat{Y}=f(X)$, and $\hat{Y}$ represents the softmax vector output $(\hat{y}_1,\hat{y}_2,...,\hat{y}_L)$. Let $n=n_1+n_2$, 
$\omega_1 = \{1,...,n_1\}$, $\omega_2 = \{n_1+1,...,n\}$. Define $D^{(1)}=\left\{\hat{Y}_{i}\right\}_{i=1}^{n_1}$, $D^{(2)}=\left\{\hat{Y}_{i}\right\}_{i=n_1+1}^{n}$, and denote the combined sample set as $D=\left\{\hat{Y}_{i}\right\}_{i=1}^{n}$, where
\begin{equation}
\hat{Y}_i =
\begin{cases}
f(X_i), & \mbox{$i \in \omega_1$}, \\
f(X'_{i-n_1}), & \mbox{$i \in \omega_2$}.
\end{cases}
\end{equation}

\subsection{Statistic of Half \texorpdfstring{$k$-} -farthest neighbor (Half-KFN)}\label{CT}
Our statistic represents an improvement upon the KNN statistic \cite{schilling1986mutual}. In the context of previous detection scenarios, KNN test plays a pivotal role, being not only conceptually straightforward but also possessing an analytically tractable null distribution. Consequently, with a sufficiently large sample size, the utilization of its simple and precise null distribution allows for intuitive detection, showcasing its inherent efficiency advantage.

However, to achieve a more sensitive detection for subtle covariate drift, enhancements to the KNN statistic are imperative. Naturally, for a small subset of drifted samples, their predictive accuracy is evidently diminished after undergoing BBSDs dimensionality reduction, compared to the normal samples unaffected by drift. In other words, when drifted samples are processed through an arbitrary label classifier $f$ for classification, resulting in a vector-form classification outcome, their predictive performance is highly likely to deteriorate significantly compared to the classification results of normal samples not subject to drift. In order to capture these deteriorated samples effectively, we have devised a novel statistic to detect instances where the predictive performance degrades due to drift.

In this paper, we employ the notation $||\cdot||$ to signify the $L_2$ norm, and define the $k$-th intra-class farthest neighbor of $\hat{Y}_i$ to be $\hat{Y}_j$, satisfying $||\hat{Y}_{j'} - \hat{Y}_i|| > ||\hat{Y}_j - \hat{Y}_i||$ for exactly $k-1$ indices ${j'}$ $(1 \leq {j'} \leq n, { j'} \neq i, j)$, and ensuring that the indices of the maximum elements in $\hat{Y}_i$ and $\hat{Y}_j$ are the same, which means $\hat{Y}_i$ and $\hat{Y}_j$ are in the same class. In simpler terms, the intra-class $k$ farthest neighbors of $\hat{Y}_i$ refer to the $k$ samples of the same class as $\hat{Y}_i$ that are farthest away from $\hat{Y}_i$. We define two events: $A_{i,j}^{(k)} \triangleq \{\hat{Y}_j\text{ is one of the intra-class $k$ farthest neighbors of } \hat{Y}_i \}$ and $B_{ij} \triangleq \{\hat{Y}_i \text{ is from } p \text{ and } \hat{Y}_j \text{ is from } q\}$.

% Half-KFN statistic can be described in two forms, both of which carry the same meaning. The first form is utilized throughout the document, while the second form is used to explain the rationale behind this definition:

% \textbf{Form 1}: Let $FN_i(r)$ represent the $r$-th intra-class farthest neighbor of $\hat{Y}_i$. Define that 
% \begin{equation}
%     I_i(r) =
%     \begin{cases}
%     1, &  FN_i(r)$ and $\hat{Y}_i$ belong to different sample sets, $\\
%     0, & $otherwise$.
%     \end{cases}
% \end{equation}

% The Half-KFN statistic considered when testing $H_0$ are represented as follows:

% \begin{equation}
%     T_{k,n_1} =  \frac{1}{n_1 k}\sum\limits_{i=1}^{n_1}\sum\limits_{r=1}^{k}I_i(r)
% \end{equation}

% \textbf{Form 2}: 
Two sets of events are represented by indicator functions as follows: 
\begin{equation*}
        I(A_{i,j}^{(k)}) = \left\{\begin{array}{lcl}
		1, \text{ }A_{i,j}^{(k)} \text{ occurs }( i\neq j),\\
        % 1\leq i\neq j \leq n,\\
		0, \text{ otherwise}.
	\end{array}\right.
\end{equation*}
\begin{equation*}
	I(B_{ij}) = \left\{\begin{array}{lcl}
		1, \text{ }B_{ij}\text{ occurs},\\
        % 1\leq i\neq j \leq n,\\
		0,\text{ otherwise}.
	\end{array}\right.
 % I(\{\hat{Y}_i \hspace{0.2cm} \text{is from } p \hspace{0.2cm}\text{and}\hspace{0.2cm} \hat{Y}_j \hspace{0.2cm} \text{is from } q \}).
\end{equation*}

Thus, $I(A_{i,j}^{(k)})$ can be assembled to form the following matrix $\mathcal{A}^{(k)}$, which facilitates a more intuitive understanding of the rationale behind the Half-KFN construction:
\begin{equation}\label{halfKFN,(I(Aij))}
\resizebox{1\hsize}{!}{$
    \begin{aligned}
        &
        \begin{array}{c@{\hspace{2pt}}l}
            \left(
                \begin{array}{ccc|cc:ccc}
                0 & \cdots & I(A_{1,n_1}^{(k)}) & I(A_{1,n_1+1}^{(k)}) & \cdots & \color{orange}{\mathbf{\cdots}} & \color{orange}{\mathbf{\cdots}} & \color{orange}{\mathbf{I(A_{1,n}^{(k)})}} \\
                \vdots & \ddots & \vdots & \vdots & & & &\color{orange}{\mathbf{\vdots}} \\
                I(A_{n_1,1}^{(k)}) & \cdots & 0 & \vdots & & & & \color{orange}{\mathbf{I(A_{n_1,n}^{(k)})} }\\\cline{1-8}
                I(A_{n_1+1,1}^{(k)}) & \cdots & \cdots & 0 & & & & \color{orange}{\mathbf{I(A_{n_1+1,n}^{(k)})}}\\
                \vdots & & & & \ddots & & &\color{orange}{\mathbf{\vdots}} \\\cdashline{1-8}
                \color{orange}{\mathbf{\vdots}} & & & & &
                 \color{orange}{\mathbf{\ddots}}& &\color{orange}{\mathbf{\vdots}}\\
                \color{orange}{\mathbf{\vdots}} &  &   &  &  & & \color{orange}{\mathbf{\ddots}} &\color{orange}{\mathbf{I(A_{n-1,n}^{(k)})}} \\
                \color{orange}{\mathbf{I(A_{n,1}^{(k)})}} & \color{orange}{\mathbf{\cdots}} &  \color{orange}{\mathbf{I(A_{n,n_1}^{(k)})}} & \color{orange}{\mathbf{I(A_{n,n_1+1}^{(k)})}} & \color{orange}{\mathbf{\cdots}} & 
                \color{orange}{\mathbf{\cdots}} & \color{orange}{\mathbf{I(A_{n,n-1}^{(k)})}}&\color{orange}{\mathbf{0}}
                \end{array}
            \right)
            \begin{array}{l}
                % \left.仅表示与\right\}配对，什么都不输出
                \\
                \\
                \\
                \\
                \begin{tikzpicture}[very thick]
                                        % Gray rectangle 
                                        \node[rotate = 90] {$\underbrace{\rule{10mm}{0mm}}_{no-drift}$};
                                \end{tikzpicture}
                \\
                \begin{tikzpicture}[very thick]
                                        % Gray rectangle 
                                        \node[rotate = 90] {$\underbrace{\rule{12mm}{0mm}}_{drift}$};
                                \end{tikzpicture}
            \end{array}
            \begin{array}{l}
                % \left.仅表示与\right\}配对，什么都不输出
                \begin{tikzpicture}[very thick]
                                        % Gray rectangle 
                                        \node[rotate = 90] {$\underbrace{\rule{15mm}{0mm}}_{source}$};
                                \end{tikzpicture}
                \\
                \begin{tikzpicture}[very thick]
                                        % Gray rectangle 
                                        \node[rotate = 90] {$\underbrace{\rule{25mm}{0mm}}_{target}$};
                                \end{tikzpicture}
            \end{array}
            \\\hspace{60pt}
            \begin{array}{cc}
                \underbrace{\rule{25mm}{0mm}}_{no-drift} &
                \underbrace{\rule{38mm}{0mm}}_{drift}
            \end{array}
            \\\hspace{-60pt}
            \begin{array}{cc}
                \underbrace{\rule{45mm}{0mm}}_{source} &
                \underbrace{\rule{70mm}{0mm}}_{target}
            \end{array}
        \end{array}
    \end{aligned}
$}
\end{equation}

It should be noted that this matrix is not symmetric. For each of the $n$ samples, the positions of the $k$ farthest neighbors belonging to the same class are recorded according to rows. The values of these $k$ positions are set to $1$, while the remaining $n-k$ positions in that row are assigned $0$. 

Similarly, $I({B_{ij}})$ can be organized into the following block matrix ${\mathcal{B}}$, within which we primarily focus on the upper-right sub-block:

\begin{equation}\label{halfKFN,(I(Bij))}
\resizebox{0.43\hsize}{!}{$
    \begin{aligned}
        &
        \begin{array}{c@{\hspace{2pt}}l}
            \left(
                \begin{array}{ccc|ccc}
                0 & \cdots & 0 & 1 &\cdots& 1 \\
                \vdots & \ddots & \vdots & \vdots & \ddots & \vdots \\
                0 & \cdots & 0 & 1 & \cdots & 1  \\\cline{1-6}
                0 & \cdots & 0 & 0 &\cdots& 0 \\
                \vdots & \ddots & \vdots & \vdots & \ddots & \vdots \\
                0 & \cdots & 0 & 0 & \cdots & 0  
                \end{array}
            \right)
            \begin{array}{l}
                \begin{tikzpicture}[very thick]
                                        % Gray rectangle 
                                        ;
                                        \end{tikzpicture}
                \\
                \begin{tikzpicture}[very thick]
                                        ;
                                \end{tikzpicture}
            \end{array}
            \begin{array}{l}
                % \left.仅表示与\right\}配对，什么都不输出
                \begin{tikzpicture}[very thick]
                                        % Gray rectangle 
                                        \node[rotate = 90] {$\underbrace{\rule{15mm}{0mm}}_{source}$};
                                \end{tikzpicture}
                \\
                \begin{tikzpicture}[very thick]
                                        % Gray rectangle 
                                        \node[rotate = 90] {$\underbrace{\rule{15mm}{0mm}}_{target}$};
                                \end{tikzpicture}
            \end{array}
            \\\hspace{-40pt}
            \begin{array}{cc}
                \underbrace{\rule{15mm}{0mm}}_{source} &
                \underbrace{\rule{15mm}{0mm}}_{target}
            \end{array}
        \end{array}
    \end{aligned}
$}
\end{equation}

The Half-KFN statistic $T_{k,n_1}$ can be defined in terms of these events as follows:
\begin{equation} \label{eq:half-kfn1}
    T_{k,n_1} \triangleq \frac{1}{n_1k} \mathcal{A}^{(k)} * {\mathcal{B}}=\frac{1}{n_1k}\sum\limits_{i=1}^{n_1}\sum\limits_{j=n_1+1}^{n}I({A}_{ij}^{(k)}),
\end{equation}
which can be expressed by Hadamard product of the above two matrices. The larger the difference between the two distributions $p$ and $q$, the larger the statistic $T_{k,n_1}$ will be.

Next, we describe why $T_{k,n_1}$ is constructed in this way: 

\textbf{Why should be intra-class?} The selected $f$ is usually in the neural network classification model which generates an $L$-dimensional vector for each sample, representing a common softmax output result. However, when calculating the distances between softmax output results of each sample, most of the farthest neighbors selected belong to different classes, causing confusion in drift test. What we aim for is to accurately cover drifted samples, thereby improving the accuracy of drift sample recognition. As a result, our proposed solution is to compute the distance matrix by directly calculating the distances only between samples of the same class (considered as the same class if the indices of the maximum element in the $L$-dimensional vector are the same).

\textbf{What is Half-KFN?} We define Half-KFN, a statistic characterizing the difference between $p$ and $q$, as the proportion of intra-class $k$ farthest neighbors in the upper right of $\mathcal{A}^{(k)}$ delimited by solid lines. In order to provide a better explanation, we can further divide the samples in $q$ into drifting samples and non-drifting samples. It should be noted that the actual drift proportion is unknown. In fact, there is a high probability that the intra-class $k$ farthest neighbors are located within the $5$ bold colored matrices shown in $\mathcal{A}^{(k)}$. This is because after dimensionality reduction, the intra-class $k$ farthest neighbors of the samples in $p$ are highly concentrated in the drifting samples of $q$ (corresponding to the upper right part of $\mathcal{A}^{(k)})$. However, in $q$, due to the presence of drifting samples, its intra-class $k$ farthest neighbors could be distributed in both $p$ and $q$ (corresponding to the lower left and lower right parts of $\mathcal{A}^{(k)})$, weakening the effectiveness of the statistic measuring the intra-class $k$ farthest neighbors of the samples in $q$. This inspires us that focusing solely on the upper right part of $\mathcal{A}^{(k)}$ could well be more effective and provide stronger evidence against the null hypothesis.

Similar to $k$-nearest neighbor detection, here we also aim to reject $H_0$ for the $T_{k,n_1}$ statistic displaying large values. In cases of limited sample sizes, permutation tests are typically employed to determine percentiles based on a given confidence level, thus deciding whether to reject $H_0$. Although this method can be effective with small sample sizes, its application becomes increasingly demanding in terms of computational time and resources as the sample size grows larger. So, it becomes evident that establishing the overall distribution of the statistic is a more expedient and efficient approach.

\begin{algorithm}[H]
    \caption{Half-KFN statistic algorithm}
    \label{alg:Half-KFN statistic}
    \renewcommand{\algorithmicrequire}{\textbf{Input:}}
    \renewcommand{\algorithmicensure}{\textbf{Output:}}
    \begin{algorithmic}[1]
         \small{\REQUIRE Samples $\left\{X_{1},...,X_{n_1}\right\}$ from source distribution; samples $\left\{X'_{1},...,X'_{n_2}\right\}$ from target distribution; a trained soft-classifier $f: \mathbb{R}^{C} \rightarrow \mathbb{R}^{L}$; the number of farthest neighbors $k$.  %%input
        \ENSURE \textit{Statistic of Half-KFN}
        \STATE $D^{(1)}=\left\{\hat{Y}_{i}\right\}_{i=1}^{n_1}$ $\leftarrow$ $\left\{f(X_{i})\right\}_{i=1}^{n_1}$, 
        $D^{(2)}=\left\{\hat{Y}_{i}\right\}_{i=n_1+1}^{n}$ $\leftarrow$ $\left\{f(X'_{i-n_1})\right\}_{i={n_1+1}}^{n}$.
        \STATE  Concating $D^{(1)}$ and $D^{(2)}$ to form $D =\left\{\hat{Y}_{i}\right\}_{i=1}^{n} $
        \FOR{$i=1$ to $n_1$}
            \FOR{$j=1$ to $n$}
                \IF{$\textit{Index of }max(\hat{Y}_i) == \textit{Index of }max(\hat{Y}_j)$} 
                    \STATE Calculate euclidean distance $d_{ij}$
                \ELSE
                    \STATE $d_{ij}=0$
                \ENDIF
            \ENDFOR
            \FOR{$j=1$ to $n$}
                % \STATE $A_{i,j}^{(k)}\triangleq \{\hat{Y}_{j} \hspace{0.2cm}\text{is the \textbf{intra-class} $k$-\textbf{farthest} neighbor of}\hspace{0.2cm} \hat{Y}_{i}\}$
                \IF{$d_{ij} == $ the maximum $k$ values in set $\left\{d_{ij}\right\}_{j=1}^{n}$} 
                    \STATE $I(A_{i,j}^{(k)}) = 1$
                \ELSE
                    \STATE $I(A_{i,j}^{(k)}) = 0$
                \ENDIF
            \ENDFOR
        \ENDFOR
        \STATE $T_{k,n_1}\leftarrow \frac{1}{n_1k}\sum\limits_{i=1}^{n_1}\sum\limits_{j=n_1+1}^{n}I(A_{i,j}^{(k)})$
        \RETURN $T_{k,n_1}$}
    \end{algorithmic}
\end{algorithm}

In scenarios with sufficiently large sample sizes, we can establish the distribution function or an approximate distribution for the $T_{k,n_1}$ statistic. By leveraging known properties of probability distributions, we can make statistical inferences. This method helps avoid the need for extensive permutations, thereby saving time and computational resources.

\subsection{Permutation Test}
Once we possess a statistical measure capable of quantifying the extent of covariate drift, we can utilize hypothesis testing to formally examine the null hypothesis. It is crucial to recognize that $Y$ is a continuous probability vector, as this is an essential requirement for the algorithm to function properly.

Consider conducting a permutation test with $P$ permutations. In each permutation $t \in \{1,...,P\}$, we define the two datasets, denoted as $D^{(1,t)}=\left\{\hat{Y}_{i}^{(t)}\right\}_{i=1}^{n_1}$ and $D^{(2,t)}=\left\{\hat{Y}_{i}^{(t)}\right\}_{i=n_1+1}^{n}$, which are randomly split by $D$. It is necessary to ensure that the sample sizes of the two sets are consistent with the sizes of $D^{(1)}$ and $D^{(2)}$, respectively.

Moreover, to guarantee the absence of repeated values within the set $\left\{T_{k,n_1}^{(t)}\right\}_{t=1}^{P}$, we add minor Gaussian noise perturbations centered around zero with a variance of $\sigma_{noise}^2$, to $T_{k,n_1}$ and $T_{k,n_1}^{(t)}$, for each permutation. That is 
\begin{equation}
    T = T_{k,n_1} + N(0,\sigma_{noise}^2),
\end{equation}
\begin{equation}
    T^{(t)} = T_{k,n_1}^{(t)} + N(0,\sigma_{noise}^2).
\end{equation}

Then, we evaluate each hypothesis of interest by calculating a $p$-value in the following form:
\begin{equation} \label{eq:Ln3}
	p=\frac{\sum\limits_{t=1}^P(I(T\leq T^{(t)}))}{P}.
\end{equation}

We fix a predetermined significance level $\alpha \in (0, 1)$. After calculating the particular $p$-value pertaining to a given null hypothesis, reject it if  $p\leq\alpha$.

\begin{proposition}
\label{proposition:1}
    For every $\alpha \in (0,1)$, under $H_0$, 
    \begin{equation}
    \mathbb{P}(p\leq \alpha)\leq \alpha.
    \end{equation}
\end{proposition}

The above proposition \ref{proposition:1} serves as evidence to support the validity of such tests in controlling the Type-I error rate. By doing so, these tests effectively manage and minimize false alarms, ensuring accurate and reliable results. The proof (see Appendix \ref{Proof1}) employed the theory of permutation tests (refer to Lehmann et al. \cite{lehmann2006testing} for details).

\begin{algorithm}[ht]
    \caption{Permutation Test of Half-KFN}
    \label{alg:Permutation Test}
    \renewcommand{\algorithmicrequire}{\textbf{Input:}}
    \renewcommand{\algorithmicensure}{\textbf{Output:}}
    \begin{algorithmic}[1]
         \small{\REQUIRE Samples $\left\{X_{1},...,X_{n_1}\right\}$ from source distribution; samples $\left\{X'_{1},...,X'_{n_2}\right\}$ from target distribution; a trained soft-classifier $f: \mathbb{R}^{C} \rightarrow \mathbb{R}^{L}$; the number of farthest neighbors $k$; significance level $\alpha$; the variance of Gaussian noise $\sigma_{noise}^2$.  %%input
        \ENSURE \textit{decision} (\textit{drift} or \textit{no-drift}?) %%output
        \STATE Compute Half-KFN statistic $T_{k,n_1}$ through Algorithm \ref{alg:Half-KFN statistic}
        \STATE Add Gaussion noise: $T = T_{k,n_1} + N(0,\sigma_{noise}^2)$
        \FOR{$t=1$ to $P$}
            \STATE $(D^{(1,t)},D^{(2,t)}) = \left(\left\{\hat{Y}_{i}^{(t)}\right\}_{i=1}^{n_1},\left\{\hat{Y}_{i}^{(t)}\right\}_{i=n_1+1}^{n} \right) \leftarrow$random split of $\left\{\hat{Y}_{i}\right\}_{i=1}^{n}$
            \STATE Execute step $3$ to step $20$ of Algorithm \ref{alg:Half-KFN statistic} and compute statistic of Half-KFN$_t$: $T_{k,n_1}^{(t)}$
            \STATE Add Gaussion noises: $T^{(t)} = T_{k,n_1}^{(t)} + N(0,\sigma_{noise}^2)$
            % \IF{Half-KFN_t>Half-KFN} 
            %     \STATE $I(A_{i,l}^{(k)}) = 1$
            % \ELSE
            %     \STATE $I(A_{i,l}^{(k)}) = 0$
            % \ENDIF
        \ENDFOR
        \IF{$\frac{\sum\limits_{t=1}^P(I(T\leq T^{(t)}))}{P}<\alpha$} 
            \STATE \textit{decision} $\leftarrow$ \textit{drift}
        \ELSE
            \STATE \textit{decision} $\leftarrow$ \textit{no-drift}
        \ENDIF
        \RETURN \textit{decision}}
    \end{algorithmic}
\end{algorithm}

\subsection{Bootstrap Hypothesis Test}
\label{Bootstrap Hypothesis Test}
In the case of a sufficiently large sample size, we can compute the expectation and variance of the $T_{k,n_1}$ statistic. Although the exact distribution of $T_{k,n_1}$ may be unknown, the Central Limit Theorem informs us that the distribution of the means of $T_{k,n_1}$ approximates a normal distribution when the sample size is sufficiently large. By computing its expectation and variance, we can utilize the properties of the normal distribution for statistical inference, thereby facilitating the detection task in the context of large sample sizes.

To perform the large sample distribution test based on $T_{k,n_1}$, bootstrap is performed in populations $p$ and $q$ for a specific number of resampling operations (assumed to be $M$ times), ensuring that the sample size remains constant for each resampling operation. This process generates $M$ sets of random samples, and each set of samples undergoes the aforementioned calculations to obtain the statistic $T_{k,n_1,m}$, where $m = 1,...,M$. Subsequently, the average of these $M$ statistics, $\overline{T_{k,n_1}} = \frac{\sum\limits_{m=1}^{M}T_{k,n_1,m}}{M}$, corresponding to the $M$ sets of samples, is computed. 

\begin{thm}\label{thm3}
Let $FN_i(r)$ represent the $r$-th intra-class farthest neighbor of $\hat{Y}_i$. Define that 
\begin{equation}
    I_i(r) =
    \begin{cases}
    1, &  FN_i(r)$ and $\hat{Y}_i$ belong to different sample sets, $\\
    0, & $otherwise$.
    \end{cases}
\end{equation}

The Half-KFN statistic considered when testing $H_0$ are represented as follows:

\begin{equation}\label{eq:half-kfn2}
    T_{k,n_1} =  \frac{1}{n_1 k}\sum\limits_{i=1}^{n_1}\sum\limits_{r=1}^{k}I_i(r).
\end{equation}
\end{thm}
An alternative articulation of the Half-KFN statistic shown by Equation \ref{eq:half-kfn2} is synonymous with Equation \ref{eq:half-kfn1}. The proof of Theorem \ref{thm3} is given in the Appendix \ref{Proof3}.

Assume that as $n_1$ and $n_2$ approach infinity, we have $\lim\limits_{n \to \infty} \frac{n_i}{n} = \lambda_i$, where $i = 1, 2$. In this context, we can consider the following two probabilities:
\begin{enumerate}[(1)]
    \item $p_1(r,s) = P_{H_0}(FN_i(r)=\hat{Y}_j, FN_j(s)=\hat{Y}_i)$,
    \item $p_2(r,s) = P_{H_0}(FN_i(r)=FN_j(s))$.
\end{enumerate}

For any given $\hat{Y}_i$ and $\hat{Y}_j$, with $s$ and $r$ specified, $p_1(r,s)$ and $p_2(r,s)$ respectively represent the probabilities that they are mutually farthest neighbors and share a farthest neighbor.

Note that if the sample size $n$ is large, rather than tests based on the permutation principle, one can utilize tests based on the asymptotic null distributions of $\overline{T_{k,n_1}}$, as outlined in the following theorem.

\begin{thm}\label{thm1}
     If $n$ grows to infinity in such a way that $n_1/n \rightarrow \lambda_1, n_2/n \rightarrow \lambda_2$ for $\lambda_1, \lambda_2 \in (0, 1)$, then for any fixed dimension $d$ and any fixed $k$, $\frac{
     \overline{T_{k,n_1}} - \mu}{\sqrt{\sigma^2/M}}$ asymptotically follows a standard normal distribution under $H_0$, where
    \begin{equation}
        \mu = \lim\limits_{n \to \infty}E_{H_0}(T_{k,n_1,m}) = \lim\limits_{n \to \infty} \frac{n_2}{n-1} = \lambda_2,
    \end{equation}
    \begin{equation}
    \begin{aligned}
        \sigma^2 &= \lim\limits_{n \to \infty}Var_{H_0}(T_{k,n_1,m}) \\
        &= \lambda_2^2\frac{\sum\limits_{r=1}^{k}\sum\limits_{s=1}^{k}p_1(r,s)}{k^2}+\lambda_1 \lambda_2 \frac{\sum\limits_{r=1}^{k}\sum\limits_{s=1}^{k}p_2(r,s)}{k^2} \\
        % &+\frac{n_2(nn_1-2n+1+n_2-knn_1-kn_1n_2+3kn_1+2kn_2-2k)}{kn_1(n-1)^2(n-2)}.
    \end{aligned}
    \end{equation}
\end{thm}

The proof of Theorem \ref{thm1} is given in the Appendix \ref{Proof2}.

\begin{algorithm}[h]
    \caption{Bootstrap Hypothesis Test of Half-KFN}
    \label{alg:Bootstrap Hypothesis Test}
    \renewcommand{\algorithmicrequire}{\textbf{Input:}}
    \renewcommand{\algorithmicensure}{\textbf{Output:}}
    \begin{algorithmic}[1]
         \small{\REQUIRE Source dataset $X_{source}$; target dataset $X_{target}$; the number of resampling operation $M$;  a trained soft-classifier $f: \mathbb{R}^{C} \rightarrow \mathbb{R}^{L}$; the number of farthest neighbors $k$; significance level $\alpha$; $U \sim N(0,1)$; probability of mutually farthest neighbors $p_1(r,s)$; probability of share a farthest neighbor $p_2(r,s)$.  %%input
        \ENSURE \textit{decision} (\textit{drift} or \textit{no-drift}?) %%output
        \STATE Resample $n_1$ and $n_2$ samples from the source and target datasets respectively, for $M$ times:
        \STATE $\lambda_1 \leftarrow \frac{n_1}{n}$, $\lambda_2 \leftarrow \frac{n_2}{n}$ 
        \FOR{$m=1$ to $M$}
            \STATE $\left\{X_{1,m},...,X_{n_1,m}\right\}$ $\leftarrow$random resample from $X_{source}$
            \STATE $\left\{X'_{1,m},...,X'_{n_2,m}\right\}$ $\leftarrow$random resample from $X_{target}$
            \STATE Execute Algorithm \ref{alg:Half-KFN statistic} and compute Half-KFN$_m$: $T_{k,n_1,m}$
        \ENDFOR
        \STATE $\overline{T_{k,n_1}} \leftarrow \frac{\sum\limits_{m=1}^{M}T_{k,n_1,m}}{M}$
        \STATE $\mu \leftarrow \lambda_2$
        \STATE $\sigma^2 \leftarrow \lambda_2^2\frac{\sum\limits_{r=1}^{k}\sum\limits_{s=1}^{k}p_1(r,s)}{k^2}+\lambda_1 \lambda_2 \frac{\sum\limits_{r=1}^{k}\sum\limits_{s=1}^{k}p_2(r,s)}{k^2}$
        \IF{{$P(U>\left|\frac{
     \overline{T_{k,n_1}} - \mu}{\sqrt{\sigma^2/M}}\right|)<\alpha/2$}} 
            \STATE \textit{decision} $\leftarrow$ \textit{drift}
        \ELSE
            \STATE \textit{decision} $\leftarrow$ \textit{no-drift}
        \ENDIF
        \RETURN \textit{decision}}
    \end{algorithmic}
\end{algorithm}

The subsequent task involves estimating $p_1(r,s)$ and $p_2(r,s)$. However, as $L$ and $k$ grow larger, the computational complexity associated with these estimates escalates significantly. To mitigate this complexity, we focus on the one-dimensional maximum probability value, denoted as $max(\hat{Y}_i)$, within each vector $\hat{Y}_i = (\hat{y}_{i1},\hat{y}_{i2},...,\hat{y}_{iL})$ for $i \in \{1,...,n\}$. Additionally, it is noted that in Section \ref{Dimension Reduction}, dimensionality reduction is performed on the feature data by generating a softmax vector output using a label classifier $f$. In this vector, each dimension represents the probability value for each class, with the dimension having the highest probability value typically considered as the class to which the sample belongs. This dimension is notably representative and holds significant importance in drift detection. Consequently, we will consider drift detection based on the dimension with the highest probability value.

For example, we consider the simplest case when $k=1$. As $k$ increases, $p_1(r,s)$ and $p_2(r,s)$ are estimated using the same principles and framework as described below, but the complexity of the situation and computations grows, leading to increased time costs. 

Notice that we consider the intra-class situation, for each class $l \in \{1,...,L\}$, there are $n_{(1,l)}$ samples from $q$ and $n_{(2,l)}$ samples from $p$, resulting in a total of $n_{(l)}$ samples. Since the Half-KFN statistics of each class are independent, then $Var_{H_0}(T_{k,n_1,m}) = \sum\limits_{l=1}^L Var_{H_0}(T_{k,n_{(1,l)},m})\cdot \frac{n_{(1,l)}^2}{n_{1}^2}$. The maximum and minimum values among these $n_{(l)}$ samples are recorded as shown in Figs. \ref{mainaxis} and \ref{axis} below.

\tikzset{overlap/.style={fill=yellow!30},
    block wave/.style={thick},
    function f/.style={block wave, red!50},
    function g/.style={block wave, green!50},
    convolution/.style={block wave, blue!50},
    function g position/.style={function g, dashed, semithick},
    major tick/.style={semithick},
    axis label/.style={anchor=west},
    x tick label/.style={anchor=north, minimum width=7mm},
    y tick label/.style={anchor=east},
}
\pgfkeys{/pgf/number format/.cd,fixed,precision=1}

\pgfdeclarelayer{background}
\pgfdeclarelayer{foreground}
\pgfsetlayers{background,main,foreground}

\newcommand{\mainaxis}{
    % Main axis
    $\overbrace{
    \draw (0, 0) -- (4, 0) ;

    % Small tickmarks on the x axis
    \foreach \x/\label in {0.1/\triangle,0.3/,0.8/,1.22/,1.55/,1.87/,2.18/,2.43/,2.71/,3.1/,3.31/,3.62/,3.75/,3.9/\triangle} {
        \node[x tick label] at (\x, 0) {$\label$};
        \draw (\x, 0) -- (\x, -0.8pt);
    }

    % Labels on the $x$ axis; the llap makes the label center on the
    % number without the minus.
    \foreach \x/\label in {0.1/min,3.9/max} {
        \node[x tick label] at (\x, -0.2) {$\label$};
        \draw[major tick] (\x, 0) -- (\x, -1.5pt);
    }
    \rule{80mm}{0mm}}^{n_{(l)}} $
}

\newcommand{\axis}{
    % Main axis
    $\overbrace{
    \draw (0, 0) -- (4, 0) ;
    \draw [dashed] (2, -0.5) -- (2, 0.5) ;
    % Small tickmarks on the x axis
    \foreach \x/\label in {0.1/\star,0.3/\star,0.8/\star,1.22/\star,1.55/\star,1.87/\star,2.18/\diamondsuit,2.43/\diamondsuit,2.71/\diamondsuit,3.1/\diamondsuit,3.31/\diamondsuit,3.62/\diamondsuit,3.75/\diamondsuit,3.9/\diamondsuit} {
        \node[x tick label] at (\x, 0) {$\label$};
        \draw (\x, 0) -- (\x, -0.8pt);
    }

    % Labels on the $x$ axis; the llap makes the label center on the
    % number without the minus.
    \foreach \x/\label in {0.1/min,3.9/max} {
        \node[x tick label] at (\x, -0.2) {$\label$};
        \draw[major tick] (\x, 0) -- (\x, -1.5pt);
    }
    \rule{40mm}{0mm}}^{\alpha_{(l)}}
    \overbrace{\rule{40mm}{0mm}}^{\beta_{(l)}} $
}

\newcommand{\drawg}[1]{
    \draw[function g] (#1,0) ++(-0.5, 0) -- +(0,1) -- +(1,1) -- +(1,0);
    \draw[function g position] (#1,1.4) -- (#1,\yshift);
    \node[fill=white] at (#1, 1.25) {$g$};

    % We now slightly abuse \ifdim to determine whether
    % there's an overlap.
    \ifdim#1pt>-1pt\ifdim#1pt<1.01pt
        % Draw legend.
        \fill[overlap] (-2,1.51) rectangle +(0.15,0.15);
        % The right side of f overlaps with the left side of g:
        % 'entering'
        \ifdim#1pt<0pt
            \node[anchor=west] at (-1.85, 1.575)
                {$f(\tau)g(\pgfmathprintnumber{#1} - \tau$)};
            \begin{pgfonlayer}{background}
                \fill[overlap] (-.5,0) rectangle (#1+.5,1);
            \end{pgfonlayer}
            \draw[convolution] (-1,\yshift) -- (#1, \yshift+1cm+#1 cm);
        \else
            % The left side of f overlaps with the right side of g:
            % 'leaving'
            \node[anchor=west] at (-1.85, 1.575)
                {$f(\tau)g(\pgfmathprintnumber{#1} - \tau$)};
            \begin{pgfonlayer}{background}
                \fill[overlap] (#1-.5,0) rectangle (.5,1);
            \end{pgfonlayer}
            \draw[convolution] (-1,\yshift) -- +(1, 1) --
                (#1,\yshift+1cm-#1 cm);
        \fi
    \else
        % 'g' is completely past 'f', draw the result.
        \draw[convolution] (-1,\yshift) -- +(1,1) -- +(2,0);
    \fi\fi
}

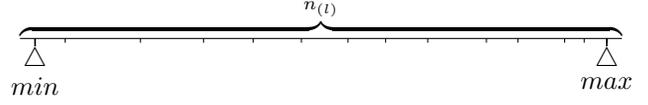
\begin{figure}[!t]
\begin{center}
    \begin{tikzpicture}[scale=2]
        \mainaxis
    \end{tikzpicture}
\end{center}
\caption{When considering the dimension $max(\hat{Y}_i)$ with $k=1$, there exists only one pair of samples that are mutual farthest neighbors, distinctly marked by triangles.}\label{mainaxis}
\end{figure}

\begin{figure}[!t]
\begin{center}
    \begin{tikzpicture}[scale=2]
        \axis
    \end{tikzpicture}
    \caption{When considering the dimension $max(\hat{Y}_i)$ with $k=1$, pairs of samples located on the same side of the dashed line $(\frac{min+max}{2})$ share a farthest neighbor.}
    \label{axis}
\end{center}
\end{figure}
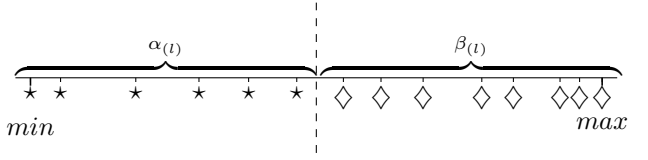
From Fig. \ref{mainaxis}, it can be observed that there is only one pair of samples that are mutually farthest neighbors (marked by triangles). Meanwhile, pairs of samples sharing a farthest neighbor can only be simultaneously distributed on one side of the dashed line in Fig. \ref{axis}. In other words, pairs of samples distributed on the left side of the dashed line (marked by stars) share a farthest neighbor, i.e, the minimum point, while pairs of samples distributed on the right side of the dashed line (marked by diamonds) also share a farthest neighbor, i.e, the maximum point.

\begin{equation}\label{p1(1,1)}
    p_1(1,1) = \lim\limits_{n \to \infty} \sum\limits_{l=1}^{L} \frac{1}{\binom{n_{(l)}}{2}} \cdot \frac{n_{(1,l)}^2}{n_{1}^2},
\end{equation}
\begin{equation}\label{p2(1,1)}
    p_2(1,1) = \lim\limits_{n \to \infty}  \sum\limits_{l=1}^{L}\frac{\binom{\alpha_{(l)}}{2}+\binom{\beta_{(l)}}{2}}{\binom{n_{(l)}}{2}} \cdot \frac{n_{(1,l)}^2}{n_{1}^2}.
\end{equation}

We can keep track of the number of points falling to the left of the dashed line as $\alpha_{(l)}$ and the number falling to the right as $\alpha_{(2,l)}$ among $n_{(l)}$ samples, where $\sum\limits_{l=1}^{L}(\alpha_{(l)} + \beta_{(l)}) = n$. This way, we can estimate $p_1(1,1)$ and $p_2(1,1)$ based on this information as shown in the formula \ref{p1(1,1)} and \ref{p2(1,1)}.

Hence, the asymptotic variance can be simplified into a more straightforward form. 
Experimental evidence suggests that this method of statistical inference is more quickly in larger sample sizes, and exhibits higher power when testing for subtle drift compared to other detection methods.

\section{Experimental evaluation}\label{section:Experimental evaluation}
Experiments in this section utilize PyTorch's built-in distribution package. To run the experiments, we used a machine running a $64$-bit operational system with $16$ GB of RAM and a $3.20$ GHz AMD Ryzen $7$ $5800$H with Radeon Graphics CPU. The source code of Half-KFN is available online\footnote{\href{https://github.com/bbwang1030/Half-KFN}{https://github.com/bbwang1030/Half-KFN}}.

\subsection{Experiments on Simulated Data}
The simulation datasets consist of training set $\{(X_i, Y_i)\}_{i=1}^{6000}$ and test set $\{(X'_j)\}_{j=1}^{3000}$. For training set, there are a total of $3$ categories, each of which contains $2000$ samples, which means 
\begin{center}
    $X_i \sim 
    \begin{cases}
        U_5(1, 2), & 1 \leq i \leq 2000, \\
        U_5(5, 6), & 2001 \leq $i$ \leq 4000,  \\
        U_5(10, 11), & 4001 \leq $i$ \leq 6000, 
    \end{cases}
    $
    $Y_i =
    \begin{cases}
        1, &  1 \leq i \leq 2000, \\
        2, &  2001 \leq i \leq 4000,  \\
        3, &  4001 \leq i \leq 6000. 
    \end{cases}
    $ 
\end{center}

\begin{figure}[htbp]
    \centering
    \scalebox{0.3}{
    \includegraphics{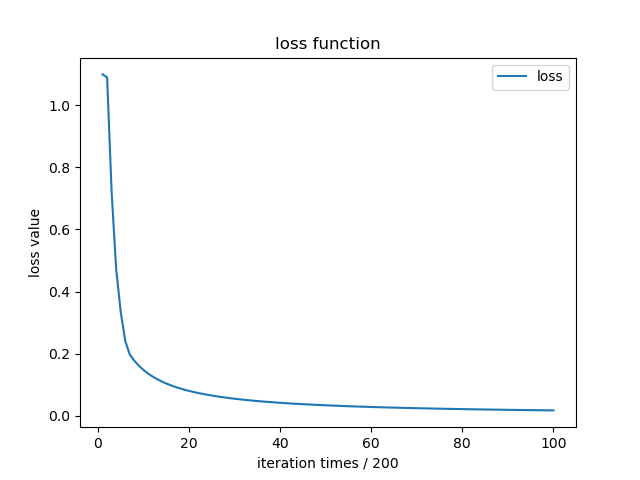}
    }
    \caption{\small The graphical representation of the loss variation during the training process of a linear classifier is obtained. The classifier parameters are estimated using gradient descent optimization. A total of $\num{2e4}$ iterations are performed, and the loss value is recorded every $250$ iterations. These recorded values are then utilized to generate a loss variation graph, enabling visualization of the optimization progress throughout the training process.}
    \label{figure:loss}
\end{figure}

\noindent The test set consists of $3000$ unlabeled samples, initially maintaining a covariate distribution consistent with the training set when no simulated drift is introduced. In the experiment, a softmax regression multiclassifier is employed as BBSDs for simplicity. During the model training process, the cross-entropy loss function is employed, and the model parameters are iteratively optimized using gradient descent. The loss function is shown in Fig. \ref{figure:loss}.

 One sample set serves as the control dataset without any modifications (analogous to the source distribution), while the other sample set serves as the simulated drift dataset (analogous to the target distribution), where a proportion of $\delta$ samples are subjected to the addition of Gaussian noise with a variance of $\sigma_{gn}$. The experiments are conducted with sample sizes of $n_1 = n_2 \in \{ 100, 200, 500, 1000\}$. The simulated drift involves adding Gaussian noise with a standard deviation of $\sigma_{gn} = 20$ to a subset of samples in the testing set, with a proportion of $\delta \in \{0, 0.01, 0.05\}$. From the results in Table \ref{table:1}, it is evident that Half-KFN exhibits the best performance among the methods evaluated.

\begin{table}[H]   
\begin{center}   
\caption{Evaluation results of of different algorithms. The results include the Type-I error (when $\delta=0$) and power (when $\delta \in \{0.01, 0.05\}$) with Gaussian noise to a subset of samples in the testing set.}  
\scriptsize
\label{table:1} 
\begin{threeparttable}
\centering
\begin{tabular}{c c c c c c c c} 
\toprule
\textbf{$\delta$} & \textbf{$n_1$} & \textbf{MMD} & \textbf{Energy} & \textbf{FR} & \textbf{KNN} & \textbf{Ours$^1$} & \textbf{Ours$^2$} \\ 
\midrule
\multirow{3}{*}{$0$} & $100$ & $0.05$& 	$0.0$8& 	$0.06$& 	$0.09$& 	$0.03$ & 	$0.04$ \\
  & $200$ & $0.03$ & 	$0.05$& 	$0.03$& 	$0.04$& 	$0.01$& 	$0.05$ \\
  & $500$ & $0.01$ & 	$0.03$& 	$0.02$& 	$0.04$&	$0.03$& 	$0.01$ \\
  & $1000$ & $0.01$ & $0.06 $&	$0.01$ &	$0.05$ &	$0.03$  &	$0.05$ \\
\midrule
\multirow{3}{*}{$0.01$} 
 &$100$& $0.05$&$0.05$&$0.06$&$0.02$&$0.12$&$\textbf{0.30}$\\
 &$200$& $0.08$&$0.09$&$0.06$&$0.03$&$0.13$&$\textbf{0.70}$\\
 &$500$& $0.05$&$0.07$&$0.04$&$0.09$&$0.37$&$\textbf{1.00}$\\
 &$1000$& $0.04$&$0.08$&$0.05$&$0.05$&$0.65$&$\textbf{1.00}$\\
\midrule
\multirow{3}{*}{$0.05$} 
&$100$&$0.07$&$0.08$&$0.06$&$0.07$&$0.41$&$\textbf{1.00}$\\
&$200$&$0.07$&$0.07$&$0.07$&$0.14$&$0.72$&$\textbf{1.00}$\\
&$500$&$0.07$&$0.19$&$0.06$&$0.11$&$0.99$&$\textbf{1.00}$\\
&$1000$&$0.08$&$0.23$&$0.09$&$0.19$&$\textbf{1.00}$&$\textbf{1.00}$\\

\bottomrule
\end{tabular} 
% \end{center} 
\begin{tablenotes}    %这行要添加， 从这开始
    \scriptsize               %这行要添加
    \item[1] Half-KFN with permutation test         %这行要添加
    \item[2] Half-KFN with bootstrap hypothesis test       %这行要添加
  \end{tablenotes}            %这行要添加
\end{threeparttable}
\end{center}
\end{table}

% \begin{figure}[h]  %[htbp]中的h是浮动的意思
%     \centering    %居中
%     \scalebox{0.5}{
%     \includegraphics{time_sample_comp_simulation.png}
%     }
%     \caption{Comparison of inference time for a single run among MMD, Energy, FR, KNN, Half-KFN with permutation test and Half-KFN with bootstrap hypothesis test on CPU. The $x$-axis denotes the sample size of $n_1=n_2=100, 200, 500, 1000$, and the $y$-axis represents the CPU inference time.}
%     \label{figure:time_simulation}
% \end{figure}

\begin{table}[H]   
\begin{center}   
\caption{\small Evaluation runtime results of different algorithms for a single run. } 
\scriptsize
\label{table:2} 
\begin{threeparttable}
\begin{tabular}{c c c c c c c} 
\toprule
\textbf{$n_1$} & \textbf{MMD} & \textbf{Energy} & \textbf{FR} & \textbf{KNN} & \textbf{Ours$^1$} & \textbf{Ours$^2$} \\ 
\midrule
$100$ &$0.23$&$0.22$	&$0.19$	&$0.19$	&$0.16$	&$\textbf{0.04} $\\
% \Xcline{1-7}{0.4pt}
$200$ &$0.50$	&$0.52$	&$0.47$	&$0.48$	&$0.44$	&$\textbf{0.08}$ \\
% \Xcline{1-7}{0.4pt}
$500$ &$3.03$	&$3.14$	&$3.01$	&$3.05$	&$2.19$	&$\textbf{0.31}$ \\
% \Xcline{1-7}{0.4pt}
$1000$ &$13.10$	&$13.79$	&$13.08$	&$13.31$	&$8.73$	&$\textbf{1.14} $\\
\bottomrule
\end{tabular} 
\begin{tablenotes}    %这行要添加， 从这开始
    \scriptsize               %这行要添加
    \item[1] Half-KFN with permutation test         %这行要添加
    \item[2] Half-KFN with bootstrap hypothesis test       %这行要添加
  \end{tablenotes}            %这行要添加
\end{threeparttable}
\end{center}   
\end{table}

In addition, we recorded the runtime of the above methods in a single run, as shown in the Table \ref{table:2}. It can be seen that with the increase in sample size, the proposed bootstrap hypothesis test of Half-KFN not only demonstrates a higher advantage in detecting power, but also has the fastest detection rate.

\subsection{Experiments on Real Data}
This section outlines the experimental framework used to evaluate the proposed approach on real data. It details the datasets, specific configurations and evaluation metrics utilized to rigorously test the approach and demonstrate its effectiveness in practical scenarios.
\subsubsection{Datasets}
We analyze two benchmark datasets, MNIST, Fashion-MNIST and CIFAR-10, to further evaluate the proposed methodology.

\textbf{MNIST} ($N_{tr} = \num{5e4}$; $N_{val} = \num{1e4}$; $N_{te} = \num{1e4}$; $D = 28 \times 28 \times 1$; $C = 10$ classes) is a handwritten digit dataset consisting of $10$ categories such as ‘0’, ‘1’, ‘2’, ‘3’, ‘4’, ‘5’, ‘6’, ‘7’, ‘8’, and ‘9’. 

\textbf{Fashion-MNIST} ($N_{tr} = \num{5e4}$; $N_{val} = \num{1e4}$; $N_{te} = \num{1e4}$; $D = 28 \times 28 \times 1$; $C = 10$ classes) is a fashion product grayscale images dataset of clothing items, such as T-shirt/top, trouser, pullover, dress, coat, sandals, shirt, sneaker, bag and ankle boots. It consists of $10$ categories, each containing $\num{6e3}$ training images and $\num{1e3}$ testing images.

\textbf{CIFAR-10} ($N_{tr} = \num{4e4}$; $N_{val} = \num{1e4}$; $N_{te} = \num{1e4}$; $D = 32 \times 32 \times 3$; $C = 10$ classes) is a natural image dataset composed of $10$ categories, such as airplane, automobile, bird, cat, deer, dog, frog, horse, ship and truck. Each category consists of $\num{5e3}$ samples in the training set and $\num{1e3}$ samples in the testing set.

\subsubsection{Dimensionality reduction mode}
\begin{figure*}[t]
\centering
\subfigure[Samples from $p$]{\includegraphics[width=2.5in]{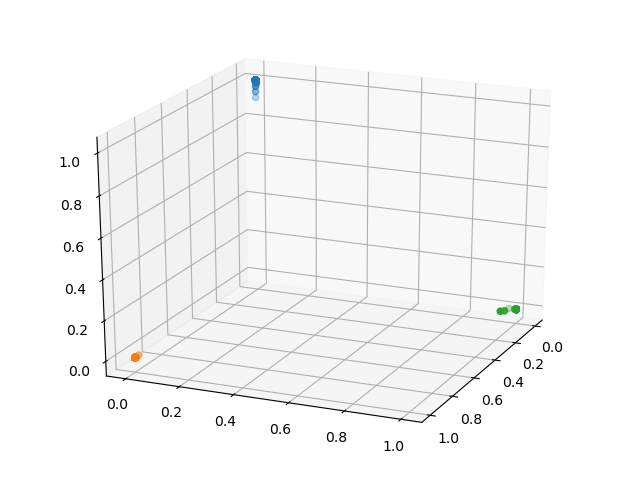}%
\label{fig_first_case}}
\hfil
\subfigure[ Samples from $q$ (with subtle drift)]{\includegraphics[width=2.5in]{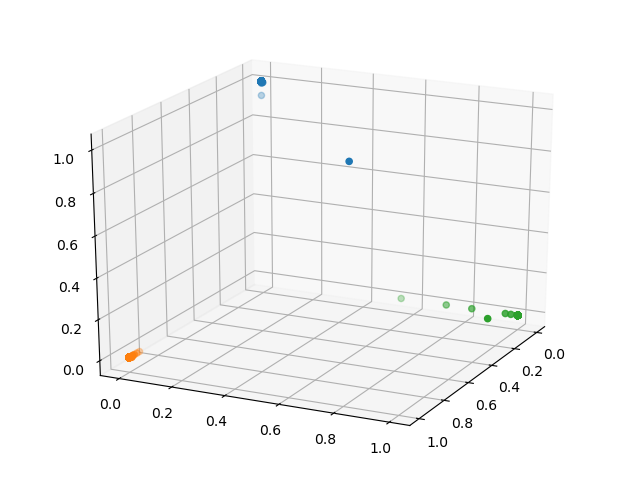}%
\label{fig_second_case}}
\caption{The visualization of the source distribution ($p$) and target distribution ($q$) of MNIST dataset after dimensionality reduction by BBSDs (which is actually 10-dimensional data, but only three dimensions are shown for visualization purposes). }
\label{fig_visual}
\end{figure*}

Subsequently, we employe a trained ResNet-$18$ as the classifier, a deep neural network with residual structure. The model employs stochastic gradient descent with momentum, processing 128 samples per batch for 200 epochs, with early stopping. The input consists of image samples from the source distribution and the target distribution. The output is the class probability vector obtained from the model for each image sample, derived through the softmax function. Fig. \ref{fig_visual} presents a visual example of dimensionality reduction using BBSDs on MNIST.

\subsubsection{Experimental parameter settings}
 The experimental simulation of each drift detection involves the process of substituting a certain proportion (denoted by $\delta \in \{0, 0.01, 0.05, 0.1\}$) of samples in the test set with drift samples. Each subplot represents the trend of Type-I error ($\delta = 0$) or detection power ($\delta \in \{0.01, 0.05, 0.1\}$) with sample sizes of $n_1=n_2 \in \{100, 200, 500, 1000\}$. In this section, we set the significant level of $\alpha=0.05$, Furthermore, we set $k=1$ in tests of KNN, SmoothKNN and Half-KFN. However, when performing bootstrap hypothesis tests, utilizing larger values of $k$ may yield benefits, albeit at the expense of a longer testing duration. Similarly, irrespective of the chosen value of $M$, our method will offer best power and maintain control the Type-I error. Nevertheless, opting for larger $M$ values will inevitably result in a longer testing time. Therefore, in these experiments we adopt $M=10$ for our bootstrap hypothesis tests. Undoubtedly, swiftly and accurately identifying drift anomalies and triggering alerts are crucial in data drift detection.
 
\subsubsection{Drift types}
Various types of drift occur, with the simulation incorporating adversarial drift, Gaussian noise drift, and image drift. 

\textbf{Adversarial drift (adv)} is characterized by the substitution of a subset of samples within the test set with adversarial samples. It aims to introduce intentionally crafted inputs. Rauber et al. \cite{rauber2020foolbox} \cite{rauber2017foolbox} launched a toolbox called Foolbox, which provides a vast array of methods for generating adversarial examples.

\textbf{Gaussian noise drift (gn)} involves the introduction of random noise, originating from a Gaussian distribution, to a specific subset of image pixels within the test set. The purpose of this manipulation is to simulate real-world scenarios where images might exhibit noise. 

\textbf{Image drift (img)} refers to the systematic augmentation of a subset of images within the test set, employing an image generator. This augmentation process comprises various transformations including rotation, horizontal and vertical translation, perspective transformation, as well as random scaling within predefined bounds. By applying this drift, the objective is to expand the diversity of encountered images during testing, which is generally related to the model's generalization ability.

\subsubsection{Execution performance} The average $p$-value is shown in Table \ref{table:pvalue1} and \ref{table:pvalue2}, and their standard deviation. The results indicate that, without drift, the average $p$-value of our test is approximately 0.5. However, in the presence of subtle drift, the average $p$-value of our test decreases, accompanied by a smaller standard deviation.

\begin{figure*}[!t]  %[htbp]中的h是浮动的意思
    \centering    %居中
    \subfigure %第一张子图
    {
        \rotatebox{90}{\scriptsize{~~~~~~~~~~~~~~MNIST}}
        \begin{minipage}[t]{0.23\textwidth}
            \centering          %子图居
            \includegraphics[width=1\textwidth]{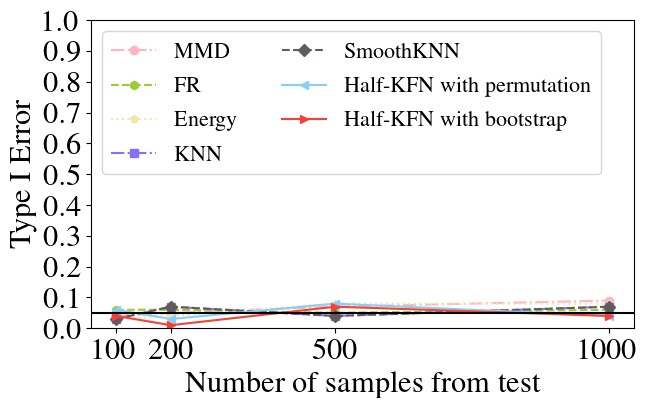}   %以行宽的0.5倍大小显示
        \end{minipage}%
    }%注意这里不能回车空行，否则两张图会上下排列，而不是并排排列
    \subfigure %第二张子图
    {
        \begin{minipage}[t]{0.23\textwidth}
            \centering          %子图居中
            \includegraphics[width=1\textwidth]{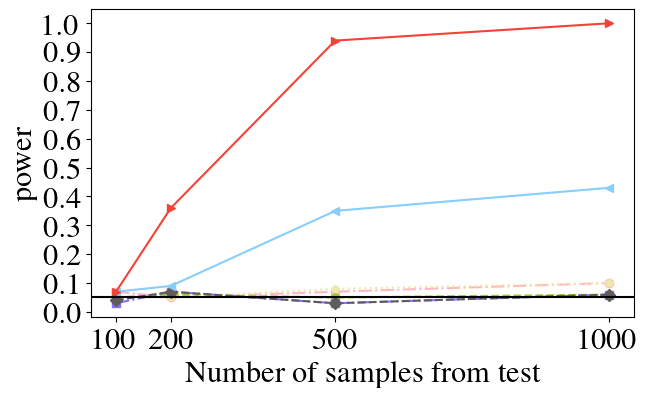}   %以行宽的0.5倍大小显示
        \end{minipage}%
    }%注意这里不能回车空行，否则两张图会上下排列，而不是并排排列
    \subfigure %第二张子图
    {
        \begin{minipage}[t]{0.23\textwidth}
            \centering          %子图居中
            \includegraphics[width=1\textwidth]{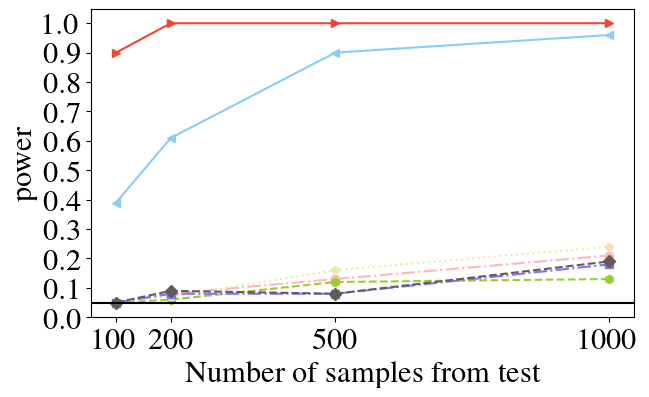}   %以行宽的0.5倍大小显示
        \end{minipage}%
    }%注意这里不能回车空行，否则两张图会上下排列，而不是并排排列
    \subfigure %第二张子图
    {
        \begin{minipage}[t]{0.23\textwidth}
            \centering          %子图居中
            \includegraphics[width=1\textwidth]{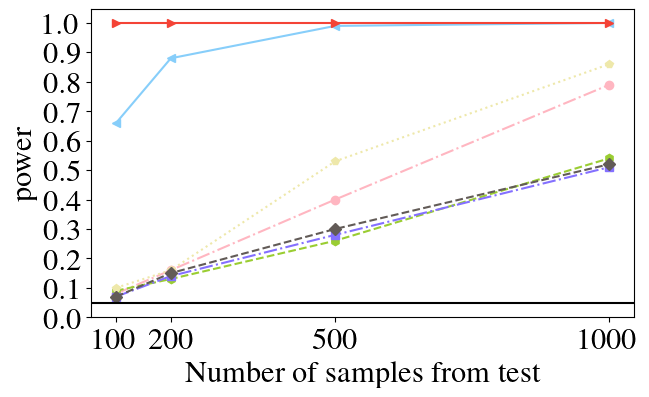}   %以行宽的0.5倍大小显示
        \end{minipage}%
    }%注意这里不能回车空行，否则两张图会上下排列，而不是并排排列

    \vspace{-3mm}
    \subfigure %第一张子图
    {
        \rotatebox{90}{\scriptsize{~~~~~~~~~Fashion-MNIST}}
        \begin{minipage}[t]{0.23\textwidth}
            \centering          %子图居
            \includegraphics[width=1\textwidth]{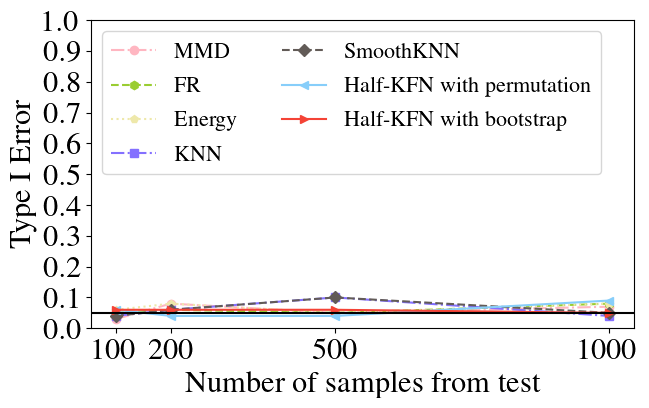}   %以行宽的0.5倍大小显示
        \end{minipage}%
    }%注意这里不能回车空行，否则两张图会上下排列，而不是并排排列
    \subfigure %第二张子图
    {
        \begin{minipage}[t]{0.23\textwidth}
            \centering          %子图居中
            \includegraphics[width=1\textwidth]{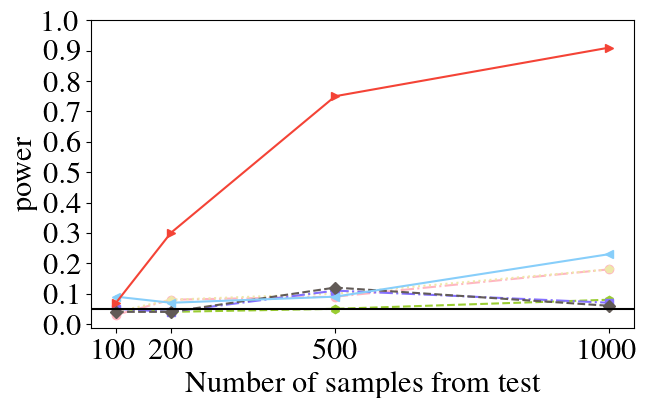}   %以行宽的0.5倍大小显示
        \end{minipage}%
    }%注意这里不能回车空行，否则两张图会上下排列，而不是并排排列
    \subfigure %第二张子图
    {
        \begin{minipage}[t]{0.23\textwidth}
            \centering          %子图居中
            \includegraphics[width=1\textwidth]{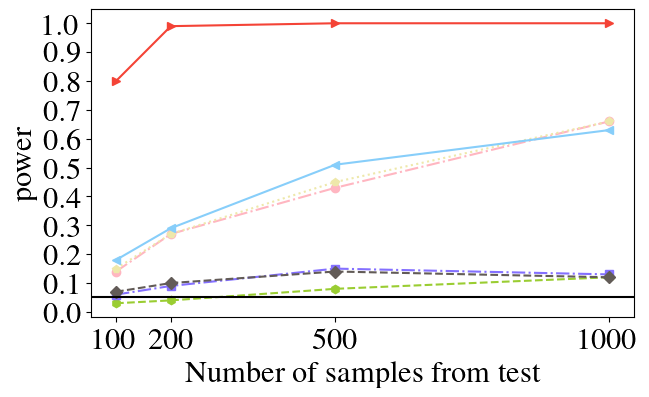}   %以行宽的0.5倍大小显示
        \end{minipage}%
    }%注意这里不能回车空行，否则两张图会上下排列，而不是并排排列
    \subfigure %第二张子图
    {
        \begin{minipage}[t]{0.23\textwidth}
            \centering          %子图居中
            \includegraphics[width=1\textwidth]{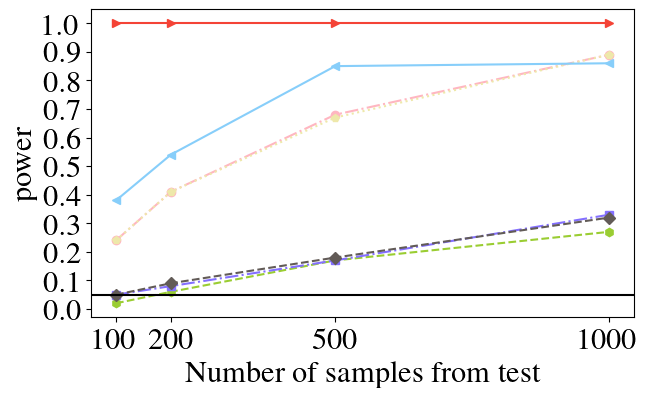}   %以行宽的0.5倍大小显示
        \end{minipage}%
    }%注意这里不能回车空行，否则两张图会上下排列，而不是并排排列
    
    \vspace{-3mm}
    \setcounter{subfigure}{0}
    \subfigure[$\delta = 0$] %第一张子图
    {
        \rotatebox{90}{\scriptsize{~~~~~~~~~~~~CIFAR-10}}
        \begin{minipage}[t]{0.23\textwidth}
            \centering          %子图居
            \includegraphics[width=1\textwidth]{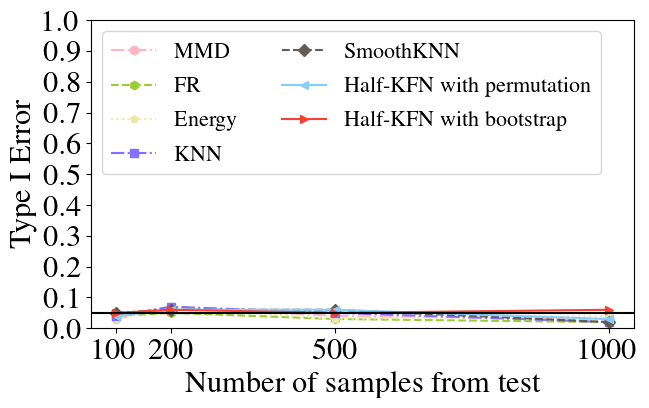}   %以行宽的0.5倍大小显示
        \end{minipage}%
    }%注意这里不能回车空行，否则两张图会上下排列，而不是并排排列
    \subfigure[$\delta = 0.01$] %第二张子图
    {
        \begin{minipage}[t]{0.23\textwidth}
            \centering          %子图居中
            \includegraphics[width=1\textwidth]{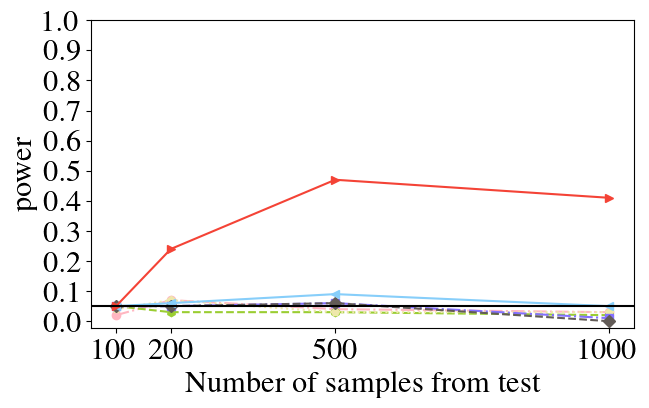}   %以行宽的0.5倍大小显示
        \end{minipage}%
    }%注意这里不能回车空行，否则两张图会上下排列，而不是并排排列
    \subfigure[$\delta = 0.05$] %第二张子图
    {
        \begin{minipage}[t]{0.23\textwidth}
            \centering          %子图居中
            \includegraphics[width=1\textwidth]{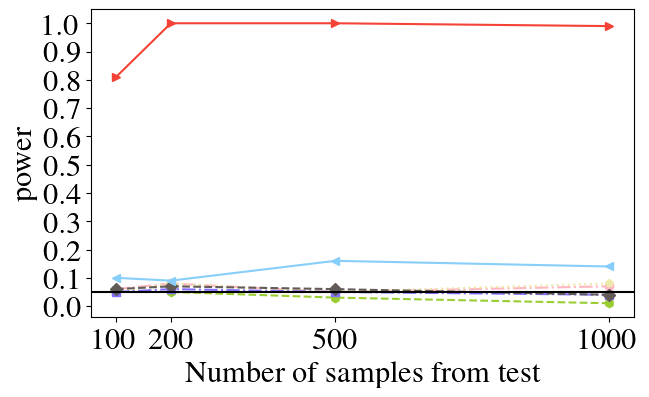}   %以行宽的0.5倍大小显示
        \end{minipage}%
    }%注意这里不能回车空行，否则两张图会上下排列，而不是并排排列
    \subfigure[$\delta = 0.1$] %第二张子图
    {
        \begin{minipage}[t]{0.23\textwidth}
            \centering          %子图居中
            \includegraphics[width=1\textwidth]{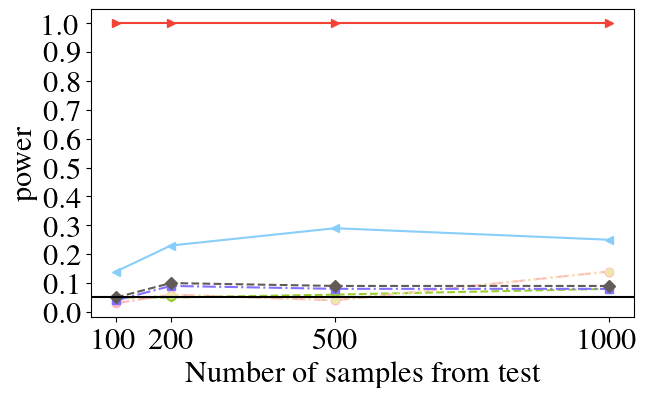}   %以行宽的0.5倍大小显示
        \end{minipage}%
    }%注意这里不能回车空行，否则两张图会上下排列，而不是并排排列
    
    \caption{\small Drift test on MNIST, Fashion-MNIST and CIFAR10 datasets with adversarial drift with parameter $\delta$. (a) Figure demonstrates that the Type-I error rates of both two datasets are near significant level $\alpha=0.05$, indicating effective control of Type-I errors. (b-d) Figures illustrate that with increasing sample size, the detection power of various two sample tests improves.} %  %大图名称
    \label{fig2}  %图片引用标记
\end{figure*}

\begin{figure*}[!t]  %[htbp]中的h是浮动的意思
    \centering    %居中
    \subfigure %第一张子图
    {
        \rotatebox{90}{\scriptsize{~~~~~~~~~~~~~~MNIST}}
        \begin{minipage}[t]{0.23\textwidth}
            \centering          %子图居
            \includegraphics[width=1\textwidth]{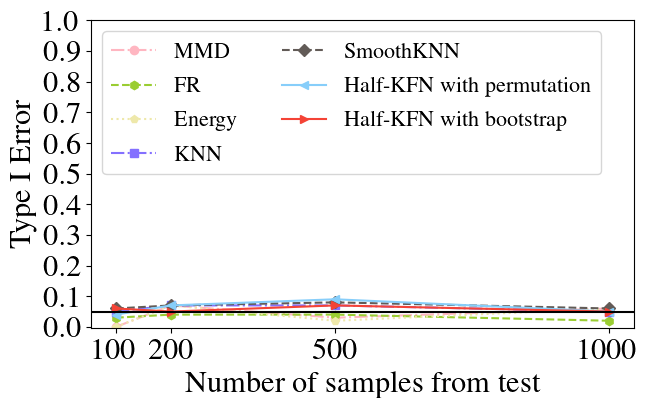}   %以行宽的0.5倍大小显示
        \end{minipage}%
    }
    \subfigure %第二张子图
    {
        \begin{minipage}[t]{0.23\textwidth}
            \centering          %子图居中
            \includegraphics[width=1\textwidth]{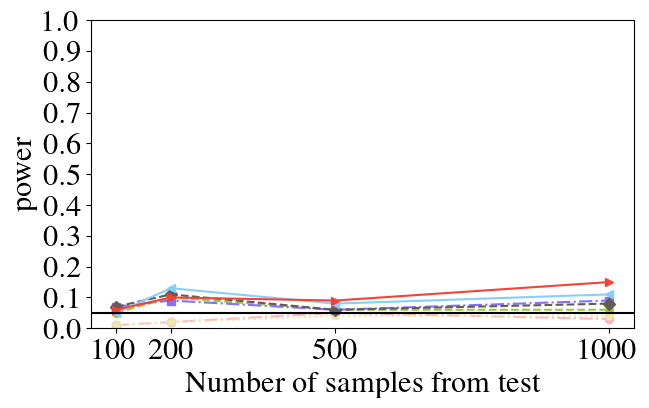}   %以行宽的0.5倍大小显示
        \end{minipage}%
    }
    \subfigure %第二张子图
    {
        \begin{minipage}[t]{0.23\textwidth}
            \centering          %子图居中
            \includegraphics[width=1\textwidth]{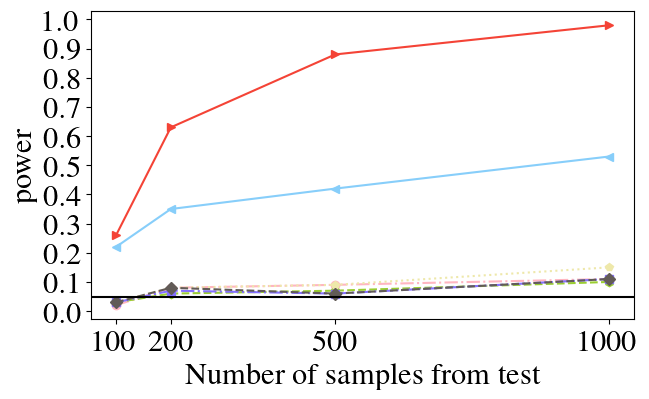}   %以行宽的0.5倍大小显示
        \end{minipage}%
    }
    \subfigure %第二张子图
    {
        \begin{minipage}[t]{0.23\textwidth}
            \centering          %子图居中
            \includegraphics[width=1\textwidth]{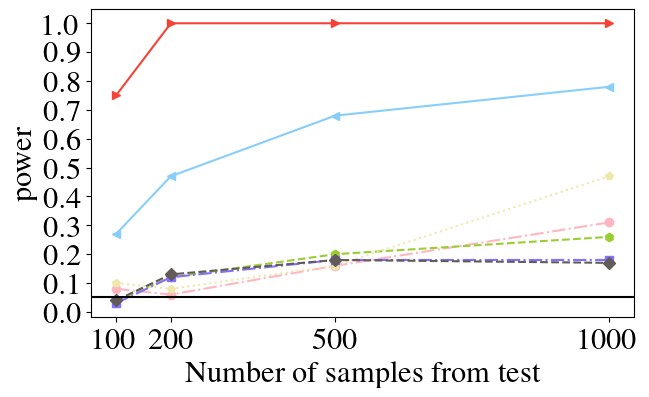}   %以行宽的0.5倍大小显示
        \end{minipage}%
    }%注意这里不能回车空行，否则两张图会上下排列，而不是并排排列

    \vspace{-3mm}
    \subfigure %第一张子图
    {
        \rotatebox{90}{\scriptsize{~~~~~~~~~Fashion-MNIST}}
        \begin{minipage}[t]{0.23\textwidth}
            \centering          %子图居
            \includegraphics[width=1\textwidth]{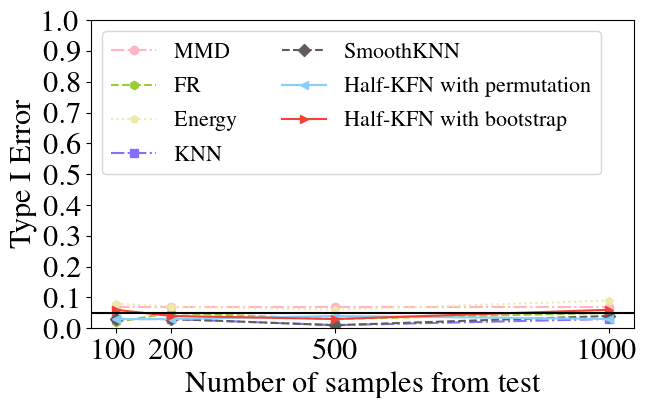}   %以行宽的0.5倍大小显示
        \end{minipage}%
    }
    \subfigure %第二张子图
    {
        \begin{minipage}[t]{0.23\textwidth}
            \centering          %子图居中
            \includegraphics[width=1\textwidth]{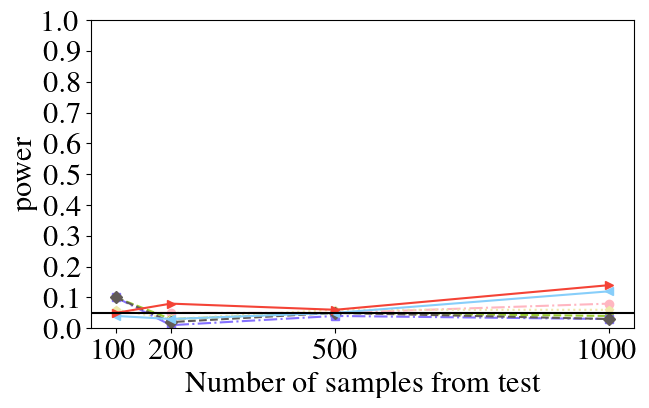}   %以行宽的0.5倍大小显示
        \end{minipage}%
    }
    \subfigure %第二张子图
    {
        \begin{minipage}[t]{0.23\textwidth}
            \centering          %子图居中
            \includegraphics[width=1\textwidth]{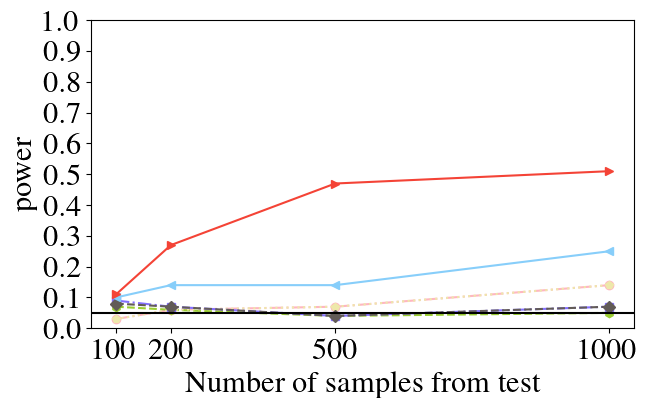}   %以行宽的0.5倍大小显示
        \end{minipage}%
    }
    \subfigure %第二张子图
    {
        \begin{minipage}[t]{0.23\textwidth}
            \centering          %子图居中
            \includegraphics[width=1\textwidth]{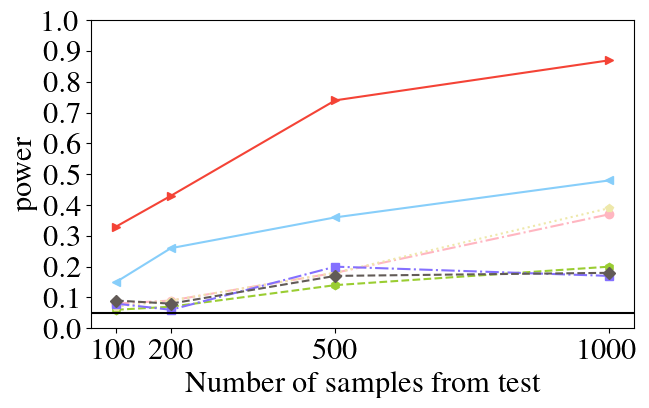}   %以行宽的0.5倍大小显示
        \end{minipage}%
    }%注意这里不能回车空行，否则两张图会上下排列，而不是并排排列
    
    \vspace{-3mm}
    \setcounter{subfigure}{0}
    \subfigure[$\delta = 0$] %第一张子图
    {
        \rotatebox{90}{\scriptsize{~~~~~~~~~~~~CIFAR-10}}
        \begin{minipage}[t]{0.23\textwidth}
            \centering          %子图居
            \includegraphics[width=1\textwidth]{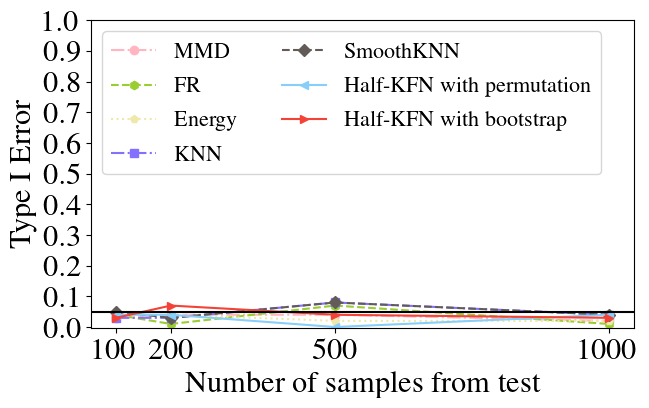}   %以行宽的0.5倍大小显示
        \end{minipage}%
    }
    \subfigure[$\delta = 0.01$] %第二张子图
    {
        \begin{minipage}[t]{0.23\textwidth}
            \centering          %子图居中
            \includegraphics[width=1\textwidth]{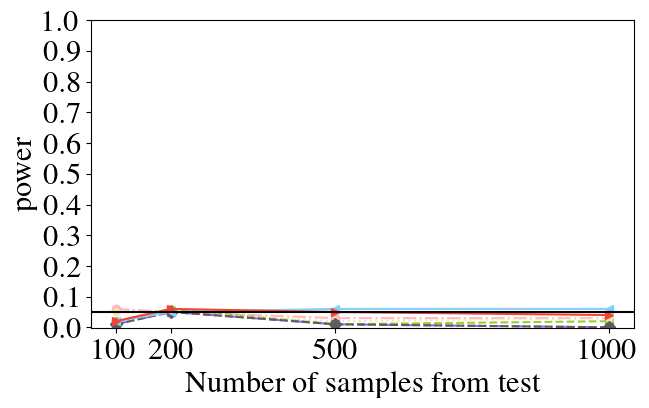}   %以行宽的0.5倍大小显示
        \end{minipage}%
    }
    \subfigure[$\delta = 0.05$] %第二张子图
    {
        \begin{minipage}[t]{0.23\textwidth}
            \centering          %子图居中
            \includegraphics[width=1\textwidth]{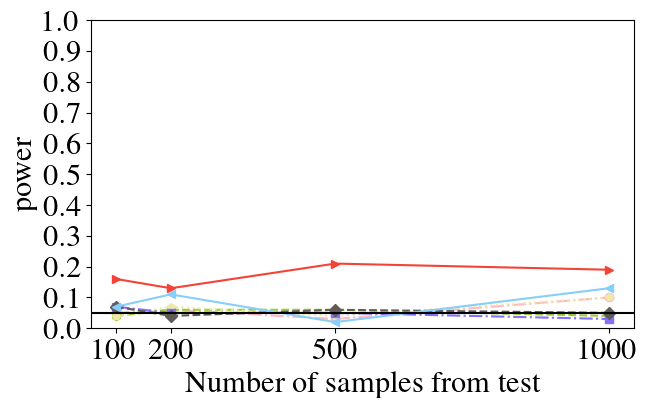}   %以行宽的0.5倍大小显示
        \end{minipage}%
    }
    \subfigure[$\delta = 0.1$] %第二张子图
    {
        \begin{minipage}[t]{0.23\textwidth}
            \centering          %子图居中
            \includegraphics[width=1\textwidth]{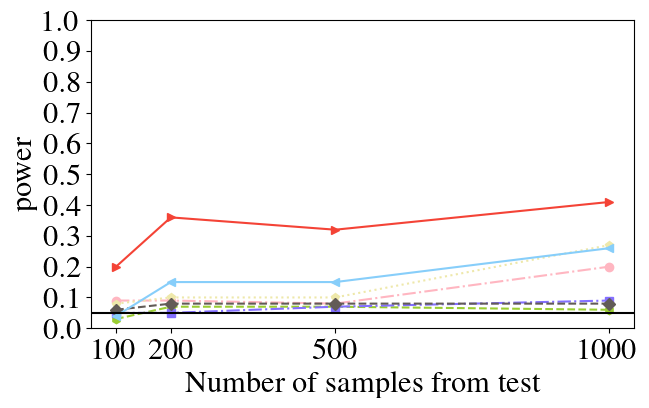}   %以行宽的0.5倍大小显示
        \end{minipage}%
    }
    
    \caption{\small Add Gaussian noise with a standard deviation of $100$ to the samples with a proportion of $\delta$ in the MNIST, Fashion-MNIST and CIFAR10 test datasets. The experimental setup and explanation are similar to Fig. \ref{fig2}.} %  %大图名称
    \label{fig:Gaussian}  %图片引用标记
\end{figure*}

\begin{figure*}[!t]  %[htbp]中的h是浮动的意思
    \centering    %居中
    \subfigure %第一张子图
    {
        \rotatebox{90}{\scriptsize{~~~~~~~~~~~~~~MNIST}}
        \begin{minipage}[t]{0.23\textwidth}
            \centering          %子图居
            \includegraphics[width=1\textwidth]{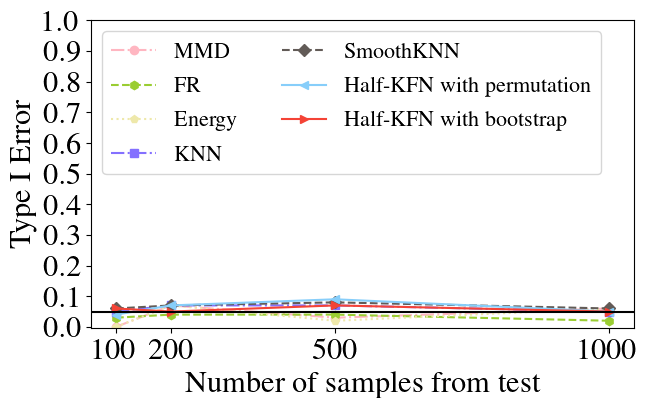}   %以行宽的0.5倍大小显示
        \end{minipage}%
    }
    \subfigure %第二张子图
    {
        \begin{minipage}[t]{0.23\textwidth}
            \centering          %子图居中
            \includegraphics[width=1\textwidth]{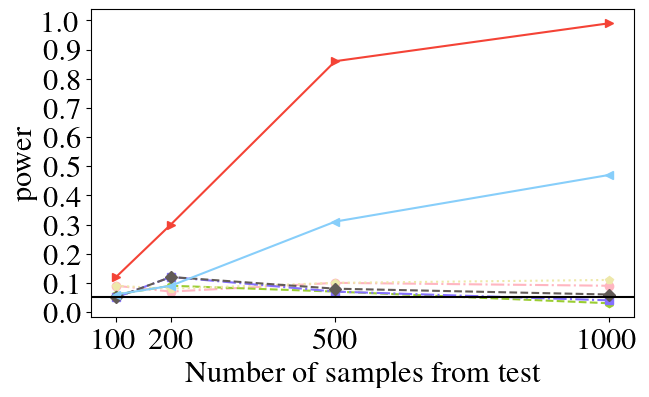}   %以行宽的0.5倍大小显示
        \end{minipage}%
    }
    \subfigure %第二张子图
    {
        \begin{minipage}[t]{0.23\textwidth}
            \centering          %子图居中
            \includegraphics[width=1\textwidth]{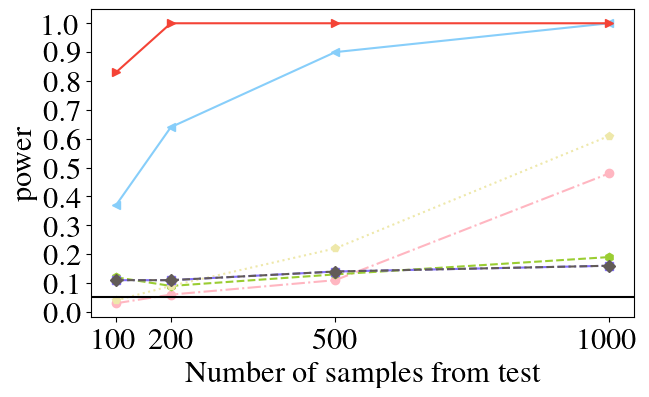}   %以行宽的0.5倍大小显示
        \end{minipage}%
    }
    \subfigure %第二张子图
    {
        \begin{minipage}[t]{0.23\textwidth}
            \centering          %子图居中
            \includegraphics[width=1\textwidth]{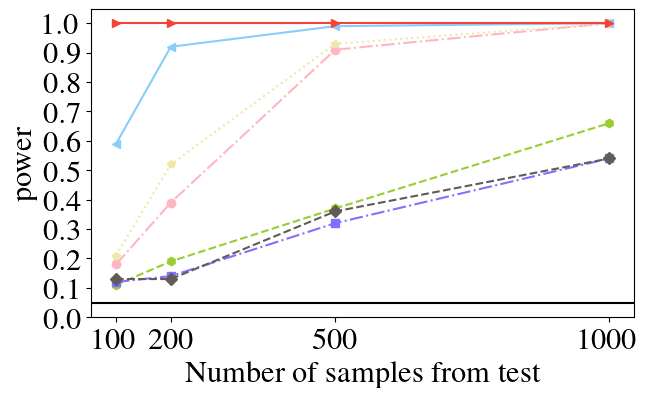}   %以行宽的0.5倍大小显示
        \end{minipage}%
    }%注意这里不能回车空行，否则两张图会上下排列，而不是并排排列
    
    \vspace{-3mm}
    \subfigure %第一张子图
    {
        \rotatebox{90}{\scriptsize{~~~~~~~~~Fashion-MNIST}}
        \begin{minipage}[t]{0.23\textwidth}
            \centering          %子图居
            \includegraphics[width=1\textwidth]{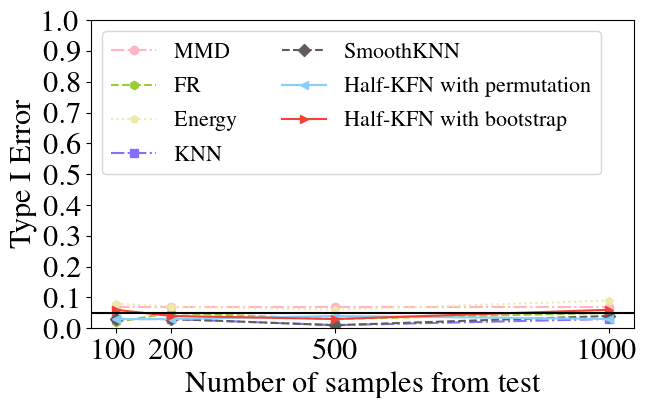}   %以行宽的0.5倍大小显示
        \end{minipage}%
    }
    \subfigure %第二张子图
    {
        \begin{minipage}[t]{0.23\textwidth}
            \centering          %子图居中
            \includegraphics[width=1\textwidth]{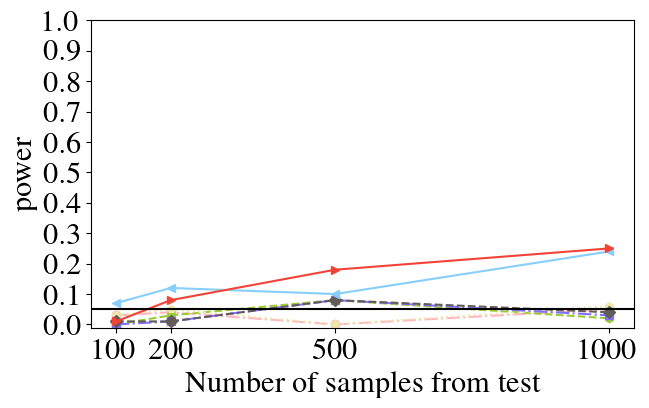}   %以行宽的0.5倍大小显示
        \end{minipage}%
    }
    \subfigure %第二张子图
    {
        \begin{minipage}[t]{0.23\textwidth}
            \centering          %子图居中
            \includegraphics[width=1\textwidth]{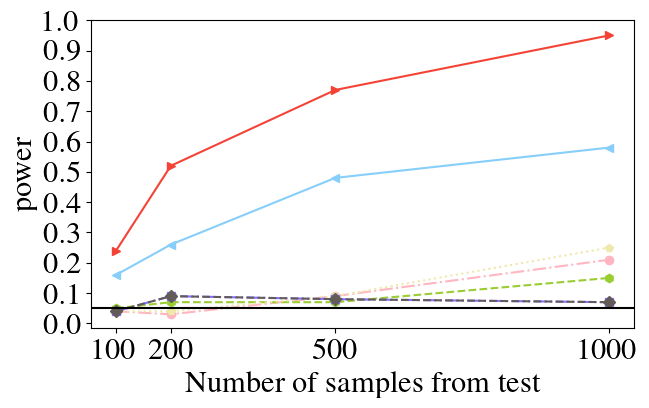}   %以行宽的0.5倍大小显示
        \end{minipage}%
    }
    \subfigure %第二张子图
    {
        \begin{minipage}[t]{0.23\textwidth}
            \centering          %子图居中
            \includegraphics[width=1\textwidth]{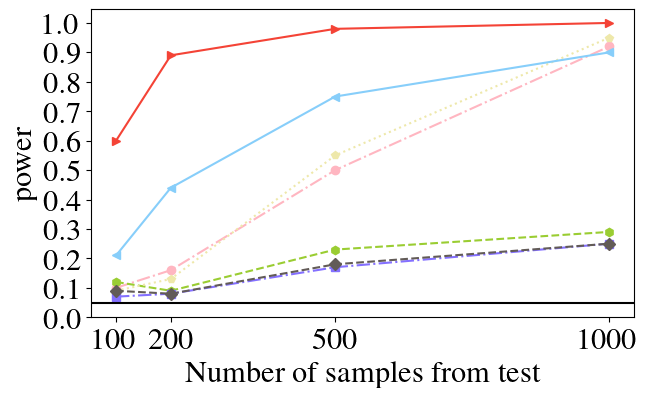}   %以行宽的0.5倍大小显示
        \end{minipage}%
    }%注意这里不能回车空行，否则两张图会上下排列，而不是并排排列
    
    \vspace{-3mm}
    \setcounter{subfigure}{0}
    \subfigure[$\delta = 0$] %第一张子图
    {
        \rotatebox{90}{\scriptsize{~~~~~~~~~~~~CIFAR-10}}
        \begin{minipage}[t]{0.23\textwidth}
            \centering          %子图居
            \includegraphics[width=1\textwidth]{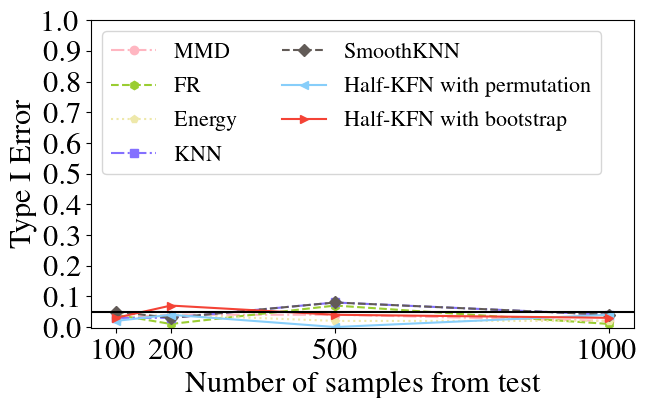}   %以行宽的0.5倍大小显示
        \end{minipage}%
    }
    \subfigure[$\delta = 0.01$] %第二张子图
    {
        \begin{minipage}[t]{0.23\textwidth}
            \centering          %子图居中
            \includegraphics[width=1\textwidth]{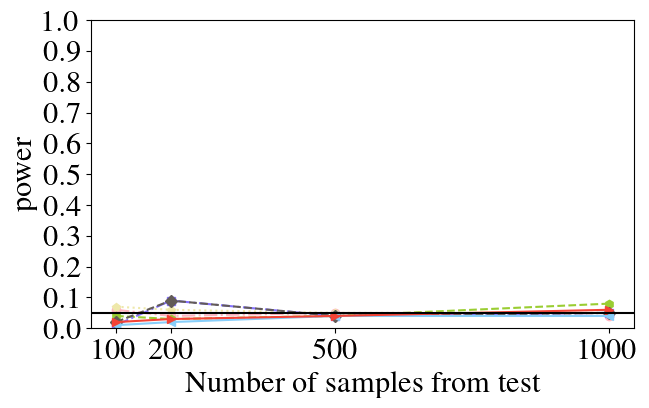}   %以行宽的0.5倍大小显示
        \end{minipage}%
    }
    \subfigure[$\delta = 0.05$] %第二张子图
    {
        \begin{minipage}[t]{0.23\textwidth}
            \centering          %子图居中
            \includegraphics[width=1\textwidth]{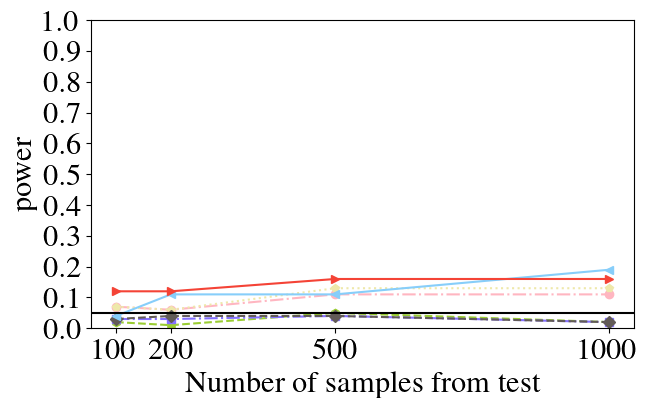}   %以行宽的0.5倍大小显示
        \end{minipage}%
    }
    \subfigure[$\delta = 0.1$] %第二张子图
    {
        \begin{minipage}[t]{0.23\textwidth}
            \centering          %子图居中
            \includegraphics[width=1\textwidth]{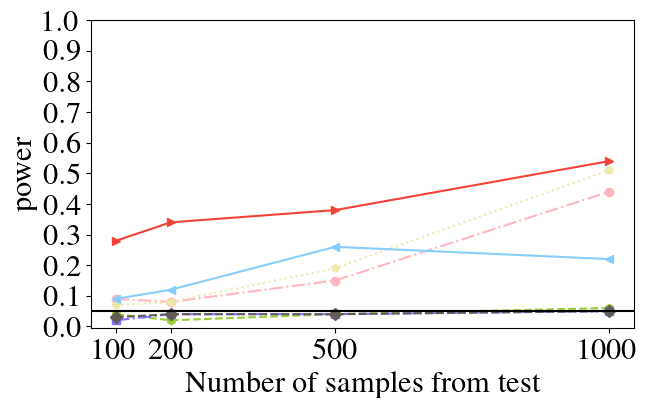}   %以行宽的0.5倍大小显示
        \end{minipage}%
    }
    \label{fig:image1}  %图片引用标记
\caption{\small Drift test on MNIST, Fashion-MNIST and CIFAR10 datasets with image drift with parameter $\delta$. Image drift refers to the systematic augmentation of a subset of images in the test set, with a proportion of $\delta$, using image generators. The experimental setup are similar to Fig. \ref{fig2}.} %  %大图名称
\label{fig:image}  %图片引用标记
\end{figure*}

We conduct comparative experiments between Half-KFN and traditional techniques, including MMD, Energy, FR, KNN, and SmoothKNN, to compare their Type-I error and power. The experimental results of these distinct drift types are depicted in Figs. \ref{fig2}, \ref{fig:Gaussian} and \ref{fig:image}, respectively. Particularly, our bootstrap hypothesis test (red line) exhibits significant advantages in terms of test performance for subtle drift. The effectiveness of the permutation test (blue line) follows.

\begin{table*}[t]   
\centering  
\caption{Drift detection results with real-world datasets, mean of $p$-value and standard deviation under $H_0$ $(\delta=0)$. Seven algorithms are run $100$ times on source and target sets of $500$ samples each, generated with different random seeds.}  
% \scriptsize
\label{table:pvalue1} 
\begin{threeparttable}
\begin{tabular}{c c c c c c c c c} 
\toprule
\scriptsize
\textbf{dataset} & \textbf{shift type} & \textbf{MMD} & \textbf{Energy} & \textbf{FR} & \textbf{KNN} &\textbf{SmoothKNN} & \textbf{Ours$^1$} & \textbf{Ours$^2$}  \\ 
\midrule
\multirow{3}{*}{MNIST} 
&adv&$0.45\pm0.28$&$0.49\pm0.30$&$0.45\pm0.28$&$0.49\pm0.28$&$0.48\pm0.28$&$0.41\pm0.28$&$0.44\pm0.27$\\
&gn&$0.50\pm0.29$&$0.48\pm0.30$&$0.50\pm0.29$&$0.45\pm0.29$&$0.44\pm0.29$&$0.47\pm0.29$&$0.52\pm0.31$\\
&img&$0.50\pm0.29$&$0.48\pm0.30$&$0.50\pm0.29$&$0.45\pm0.29$&$0.44\pm0.29$&$0.46\pm0.29$&$0.52\pm0.31$\\
\midrule
\multirow{3}{*}{Fashion-MNIST} 
&adv&$0.50\pm0.31$&$0.51\pm0.31$&$0.50\pm0.31$&$0.51\pm0.30$&$0.51\pm0.30$&$0.50\pm0.28$&$0.46\pm0.29$\\
&gn&$0.51\pm0.30$&$0.51\pm0.31$&$0.51\pm0.29$&$0.53\pm0.30$&$0.52\pm0.30$&$0.51\pm0.29$&$0.49\pm0.28$\\
&img&$0.51\pm0.30$&$0.51\pm0.31$&$0.51\pm0.29$&$0.53\pm0.30$&$0.52\pm0.30$&$0.51\pm0.29$&$0.49\pm0.28$\\
\midrule
\multirow{3}{*}{CIFAR10} 
&adv&$0.55\pm0.27$&$0.53\pm0.30$&$0.56\pm0.27$&$0.54\pm0.30$&$0.54\pm0.30$&$0.54\pm0.30$&$0.48\pm0.29$\\
&gn&$0.56\pm0.27$&$0.51\pm0.30$&$0.56\pm0.27$&$0.53\pm0.31$&$0.52\pm0.31$&$0.53\pm0.30$&$0.51\pm0.30$\\
&img&$0.56\pm0.27$&$0.51\pm0.30$&$0.56\pm0.27$&$0.53\pm0.31$&$0.52\pm0.31$&$0.53\pm0.30$&$0.51\pm0.30$\\
\midrule
  & {Avg}  & $0.52$ & 	$0.50$& 	$0.52$& 	$0.51$&	$0.50$& 	$0.50$ & $0.49$\\
\bottomrule
\end{tabular} 
% \end{center} 
\begin{tablenotes}    %这行要添加， 从这开始
    \scriptsize               %这行要添加
    \item[1] Half-KFN with permutation test         %这行要添加
    \item[2] Half-KFN with bootstrap hypothesis test       %这行要添加
  \end{tablenotes}            %这行要添加
\end{threeparttable}
\end{table*}

\begin{table*}[!t]   
\centering  
\caption{Drift detection results with real-world datasets, mean of $p$-value and standard deviation under $H_1$ $(\delta=0.05)$. Seven algorithms are run $100$ times on source and target sets of $500$ samples each, generated with different random seeds.}  
% \scriptsize
\label{table:pvalue2} 
\begin{threeparttable}
\begin{tabular}{c c c c c c c c c} 
\toprule
\scriptsize
\textbf{dataset} & \textbf{shift type} & \textbf{MMD} & \textbf{Energy} & \textbf{FR} & \textbf{KNN} &\textbf{SmoothKNN} & \textbf{Ours$^1$} & \textbf{Ours$^2$}  \\ 
\midrule
\multirow{3}{*}{MNIST} 
&adv&$0.33\pm0.26$&$0.38\pm0.29$&$0.31\pm0.25$&$0.35\pm0.26$&$0.35\pm0.26$&$0.02\pm0.05$&$\textbf{0.00}\pm\textbf{0.00}$\\
&gn&$0.46\pm0.27$&$0.43\pm0.27$&$0.45\pm0.27$&$0.44\pm0.28$&$0.44\pm0.28$&$0.15\pm0.18$&$\textbf{0.02}\pm\textbf{0.04}$\\
&img&$0.24\pm0.21$&$0.30\pm0.25$&$0.20\pm0.20$&$0.33\pm0.27$&$0.33\pm0.27$&$0.01\pm0.03$&$\textbf{0.00}\pm\textbf{0.00}$\\
\midrule
\multirow{3}{*}{Fashion-MNIST} 
&adv&$0.27\pm0.31$&$0.45\pm0.31$&$0.27\pm0.30$&$0.44\pm0.31$&$0.43\pm0.31$&$0.10\pm0.13$&$\textbf{0.00}\pm\textbf{0.00}$\\
&gn&$0.46\pm0.29$&$0.51\pm0.28$&$0.47\pm0.29$&$0.52\pm0.27$&$0.51\pm0.27$&$0.25\pm0.22$&$\textbf{0.16}\pm\textbf{0.22}$\\
&img&$0.37\pm0.25$&$0.46\pm0.29$&$0.36\pm0.25$&$0.45\pm0.27$&$0.44\pm0.27$&$0.13\pm0.17$&$\textbf{0.05}\pm\textbf{0.10}$\\
\midrule
\multirow{3}{*}{CIFAR10} 
&adv&$0.52\pm0.27$&$0.50\pm0.30$&$0.53\pm0.26$&$0.51\pm0.30$&$0.50\pm0.30$&$0.40\pm0.31$&$\textbf{0.00}\pm\textbf{0.00}$\\
&gn&$0.50\pm0.27$&$0.50\pm0.29$&$0.49\pm0.27$&$0.48\pm0.29$&$0.47\pm0.29$&$0.40\pm0.28$&$\textbf{0.32}\pm\textbf{0.32}$\\
&img&$0.45\pm0.29$&$0.50\pm0.27$&$0.45\pm0.29$&$0.50\pm0.30$&$0.49\pm0.30$&$\textbf{0.39}\pm\textbf{0.29}$&$0.39\pm0.31$\\
\midrule
  & {Avg}  & $0.40$ & 	$0.45$& 	$0.39$& 	$0.45$&	$0.44$& 	$0.21$ & $\textbf{0.10}$\\
\bottomrule
\end{tabular} 
% \end{center} 
\begin{tablenotes}    %这行要添加， 从这开始
    \scriptsize               %这行要添加
    \item[1] Half-KFN with permutation test         %这行要添加
    \item[2] Half-KFN with bootstrap hypothesis test       %这行要添加
  \end{tablenotes}            %这行要添加
\end{threeparttable}
\end{table*}

\subsubsection{Execution time}
\begin{figure}[!t]  %[htbp]中的h是浮动的意思
    \centering    %居中
    \scalebox{0.36}{
    \includegraphics{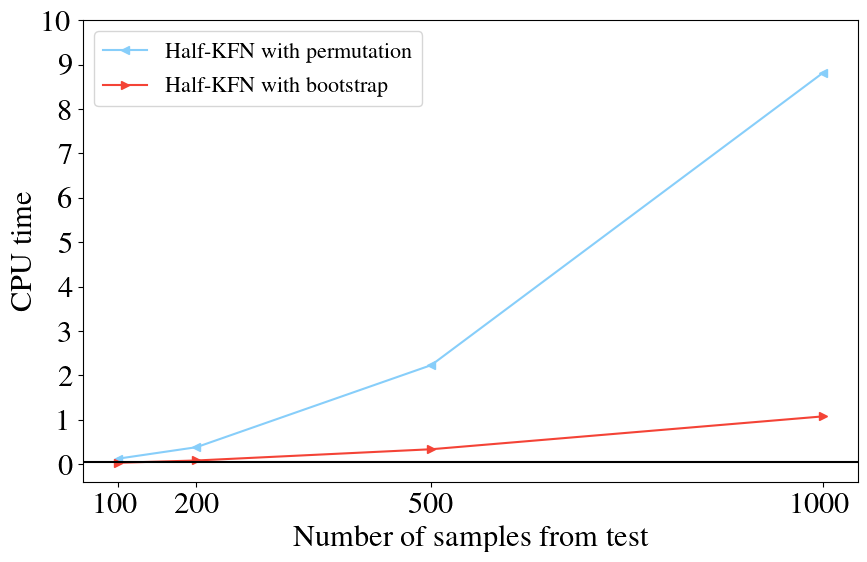}
    }
    \caption{Comparison of inference time for a single run between permutation test and bootstrap hypothesis test of Half-KFN on CPU. The $x$-axis denotes the sample size, and the $y$-axis represents the inference time.}
    \label{figure:time}
\end{figure}

Furthermore, we compared the inference time of the permutation test with the bootstrap hypothesis test in Fig. \ref{figure:time}, especially in large sample scenarios. With the gradual increase in sample size, our proposed bootstrap hypothesis test of Half-KFN demonstrates significantly shorter inference time per single run compared to the permutation test when detecting drift. This method not only exhibits higher detection power but also showcases faster detection speed, presenting a considerable advantage from various perspectives.

\section{Conclusion and Discussion}\label{section:Conclusion and Discussion}
\subsection{Conclusion}
In supervised learning, failure to promptly and accurately detect subtle covariate drift may cause a decline in model performance. To address the issue of small proportions of drift samples, this paper introduces a novel and efficient method  called Half-KFN. In addition, we calculate the expectation and asymptotic variance of Half-KFN statistic, and we propose using bootstrap to construct critical values for conducting hypothesis testing.

The following findings are observed: 

1) Compared to traditional methods, the Half-KFN method exhibits a higher power, thereby demonstrating its reliability in detecting covariate drifts

2) Heightened sensitivity in detecting small proportions of covariance shift is exhibited by the Half-KFN, utilizing both permutation test and bootstrap hypothesis test, particularly in settings with smaller sample sizes. Conversely, when dealing with larger sample sizes, the bootstrap hypothesis test of the Half-KFN proves to be not only more precise but also quicker than permutation test for detecting even the subtle covariance drifts.

3) Experiments are conducted with different types and magnitudes of drift using simulated and real-world datasets, such as MNIST, Fashion-MNIST and CIFAR10. The stability of the Half-KFN is reflected in its ability to effectively control the Type-I error under the null hypothesis, thus ensuring reliable detection.

\subsection{Discussion}
Despite the constraints imposed by the limited scope of this paper, we have not been able to exhaustively present the specific scenarios where $n_1$ and $n_2$ are unequal in our experimental results. Nevertheless, based on the theorem elaborated in this paper, it is anticipated that under such conditions, the outcomes should still align with and confirm the conclusions drawn by the theorem. As the number of distant neighbors ($k$) and bootstrap iterations ($M$) increase, the principles and framework presented in this paper enable a more efficient detection of covariate drift. However, this enhancement comes at the cost of a notable surge in computational complexity. Furthermore, when dealing with higher data dimensions ($L$), the detection process becomes more intricate, posing additional challenges. Consequently, due to space limitations, an in-depth investigation and detailed analysis of these scenarios will be undertaken in future research endeavors.

% \appendix
{\appendices
\section{Proof of Proposition \ref{proposition:1}}\label{Proof1}
\begin{proof}
For all $t \in \{1,...,P\}$, define $T = T_{k,n_1}+\epsilon$ and $T^{(t)} = T_{k,n_1}^{(t)}+\epsilon_t$, where $\epsilon$ and $\epsilon_t$'s are i.i.d from $N(0,\sigma^2)$.

Recall that 
\begin{equation}
    p=\frac{\sum\limits_{t=1}^P(I(T\leq T^{(t)}))}{P}.
\end{equation}

If $p\cdot P=a$ means there are $a$ values of $t$ that no less than $T$. If H0 in Section \ref{Problem setup} holds and the permutations are drawn uniformly from the set of all permutations of $\{1,...,P\}$, then $T,T^{(1)},...T^{(P)}$ are exchangeable. Because we add $\epsilon_t$’s in $T^{(t)}$’s, we can guarantee that all $T^{(t)}$’s are diﬀerent with probability $1$. Consequently, $U_P = p\cdot P$ is uniformly distributed in $\{1,...,P\}$. Thus,
\begin{align}
    \mathbb{P}(p\leq \alpha) &= \mathbb{P}(p\cdot P)\leq \alpha\cdot P)\\ 
    &= \mathbb{P}(U_P \leq \lfloor\alpha\cdot P\rfloor)\\ 
    &= \frac{\lfloor\alpha\cdot P\rfloor}{P}\\
    &\leq \frac{\alpha\cdot P}{P}\\
    &=\alpha
\end{align}
where $\lfloor\cdot\rfloor$ is the ﬂoor function.
\end{proof}

\section{Proof of Theorem \ref{thm3}}\label{Proof3}
\begin{proof}
    Since both Equation \ref{eq:half-kfn1} and Equation \ref{eq:half-kfn2} involve summation over $i$ from $1$ to $n_1$, we only need to prove:
    \begin{equation} \label{eq:half-kfnproof1}
        \sum\limits_{j=n_1+1}^{n}I({A}_{ij}^{(k)}) = \sum\limits_{r=1}^{k}I_i(r)
    \end{equation}
    For $i \in \{1,...,n_1\}$, so $\hat{Y}_i \in p$, then
    \scriptsize{
    \begin{equation} \label{eq:half-kfnproof2}
    \begin{aligned}
        &\sum\limits_{j=n_1+1}^{n}I({A}_{ij}^{(k)}) \\
        & = \sum\limits_{j=n_1+1}^{n}I(\{\hat{Y}_j \text{ is one of the intra-class $k$ farthest neighbors of} \hat{Y}_i\}) \\
        & = \sum\limits_{j=1}^{n}I(\{\hat{Y}_j\in q \text{ is one of the intra-class $k$ farthest neighbors of } \hat{Y}_i\}) \\
        & = \sum\limits_{j=1}^{n}\sum\limits_{r=1}^{k}I(\{\hat{Y}_j\in q \text{ is the $r$-th intra-class farthest neighbor of }\hat{Y}_i\}) \\
        & = \sum\limits_{r=1}^{k}\sum\limits_{j=1}^{n}I(\{\hat{Y}_j\in q \text{ is the $r$-th intra-class farthest neighbor of } \hat{Y}_i\}) \\
        & = \sum\limits_{r=1}^{k}I(\{\text{find the $r$-th intra-class farthest neighbor of } \hat{Y}_i, \text{ and it belongs to } q \}) \\
        & = \sum\limits_{r=1}^{k}I(\{FN_i(r) \in q \}) \\
        & = \sum\limits_{r=1}^{k}I_i(r)
    \end{aligned}
    \end{equation}
}
\end{proof}

\section{Proof of Theorem \ref{thm1}}\label{Proof2}
\begin{proof}
For $r=1,2,...,k$ and $i = 1,2,...,n$, let the $r$-th intra-class farthest neighbor of $\hat{Y}_i$ be denoted by $FN_i(r)$. Now, for $i \neq j$, let us consider the following five mutually exclusive and exhaustive probabilities:

$p_1(r,s) = P_{H_0}(FN_i(r)=\hat{Y}_j, FN_j(s)=\hat{Y}_i)$

$p_2(r,s) = P_{H_0}(FN_i(r)=FN_j(s))$

$p_3(r,s) = P_{H_0}(FN_i(r)=\hat{Y}_j, FN_j(r) \neq \hat{Y}_i)$

$p_4(r,s) = P_{H_0}(FN_i(r) \neq \hat{Y}_j, FN_j(r) = \hat{Y}_i)$

$p_5(r,s) = P_{H_0}(FN_i(r) \neq \hat{Y}_j, FN_j(r) \neq \hat{Y}_i, FN_i(r) \neq FN_j(s))$

Now, using

$p_1(r,s)=\frac{1}{n-1}P_{H_0}(FN_i(r)=\hat{Y}_j| FN_j(s)=\hat{Y}_i)$

we easily obtain $p_3(r,s) = p_4(r,s) = \frac{1}{n-1} - p_1(r,s)$ and $p_5(r,s) = \frac{n-3}{n-1}+p_1(r,s)-p_2(r,s)$

Define $I_i(r)$ as the indicator variable that takes the value $1$ if $\hat{Y}_i$ and its $r$-th farthest neighbor belong to the different sample. Now,

\begin{equation}
\begin{aligned}
T_{k,n_1} = \frac{1}{n_1 k}\sum\limits_{i=1}^{n_1}\sum\limits_{r=1}^{k}I_i(r)
\end{aligned}
\end{equation}

\begin{equation}
\begin{aligned}
E_{H_0}(T_{k,n_1}) = \frac{1}{n_1 k} n_1 k \frac{n_2}{n-1} = \frac{n_2}{n-1}
\end{aligned}
\end{equation}

\begin{equation}
\begin{aligned}
    & Var_{H_0}(T_{k,n_1}) \\
    &= \frac{1}{n_1^2k^2}Var_{H_0}\left(\sum\limits_{i=1}^{n_1}\sum\limits_{r=1}^{k}I_i(r)\right) \\
    &= \frac{1}{n_1^2k^2}\left(E_{H_0}\left(\sum\limits_{i=1}^{n_1}\sum\limits_{r=1}^{k}I_i(r)\right)^2-\left(E_{H_0}(\sum\limits_{i=1}^{n_1}\sum\limits_{r=1}^{k}I_i(r))\right)^2\right) \\
    &= \frac{1}{n_1^2k^2}\left(\sum\limits_{i=1}^{n_1}\sum\limits_{j=1}^{n_1}\sum\limits_{r=1}^{k}\sum\limits_{s=1}^{k}P(I_i(r) = I_j(s) = 1)\right)-(\frac{n_2}{n-1})^2 \\
    &=\frac{1}{n_1^2k^2}\left(\sum\limits_{i=1}^{n_1}\sum\limits_{r=1}^{k}P_{H_0}(I_i(r)=1)\right) \\
    &+\frac{1}{n_1^2k^2}\left(\sum\limits_{i=1}^{n_1}\sum\limits_{r=1}^{k}\sum\limits_{s=1,s \neq r}^{k}P(I_i(r)= I_i(s) = 1)\right) \\
    &+\frac{1}{n_1^2k^2}\left(\sum\limits_{i=1}^{n_1}\sum\limits_{j=1,j \neq i}^{n_1}\sum\limits_{r=1}^{k}\sum\limits_{s=1}^{k}P(I_i(r) = I_j(s) = 1)\right)\\
    &-(\frac{n_2}{n-1})^2
\end{aligned}
\end{equation}
% }
Note that 

\begin{equation}
\begin{aligned}
    \sum\limits_{i=1}^{n_1}\sum\limits_{r=1}^{k}P_{H_0}(I_i(r)=1) = \frac{kn_1n_2}{n-1}
\end{aligned}
\end{equation}
\begin{equation}
\begin{aligned}
    &\sum\limits_{i=1}^{n_1}\sum\limits_{r=1}^{k}\sum\limits_{s=1,s \neq r}^{k}P(I_i(r)= I_i(s) = 1)\\
    &= n_1k(k-1)\frac{n_2(n_2-1)}{(n-1)(n-2)}
\end{aligned}
\end{equation}
\begin{equation}
\begin{aligned}
    &\sum\limits_{i=1}^{n_1}\sum\limits_{j=1,j \neq i}^{n_1}\sum\limits_{r=1}^{k}\sum\limits_{s=1}^{k}P(I_i(r) = I_j(s) = 1) \\
    &= n_1(n_1-1)\sum\limits_{r=1}^{k}\sum\limits_{s=1}^{k}\beta(r,s)
\end{aligned}
\end{equation}

where 

\begin{equation}
\begin{aligned}
    \beta(r,s) = \frac{n_2}{n-2}p_2(r,s)+\frac{n_2(n_2-1)}{(n-1)(n-2)}p_5(r,s)
\end{aligned}
\end{equation}

$Var_{H_0}(T_{k,n_1})$ can be simplified to the expression:
\begin{equation}
\begin{aligned}
    &Var_{H_0}(T_{k,n_1}) \\
    &= \frac{n_2(n_2-1)(n_1-1)}{n_1(n-1)(n-2)}\frac{\sum\limits_{r=1}^{k}\sum\limits_{s=1}^{k}p_1(r,s)}{k^2}\\
    &+\frac{n_2(n_1-1)}{(n-1)(n-2)}\frac{\sum\limits_{r=1}^{k}\sum\limits_{s=1}^{k}p_2(r,s)}{k^2} \\
    &+\frac{n_2(nn_1-n-n_1-kn_1(n+n_2)+k(3n_1+2n_2-2)+1)}{kn_1(n-1)^2(n-2)}
\end{aligned}
\end{equation}

Bootstrap is performed in populations $p$ and $q$, assuming $M$ resampling operations, with each operation maintaining a consistent sample size. This process generates $M$ sets of random samples, with each set undergoing the aforementioned calculation process to obtain the statistic $T_{k,n_1,m}$, where $m = 1,...,M$. Subsequently, the average of these $M$ statistics is computed as $\overline{T_{k,n_1}} = \frac{\sum\limits_{m=1}^{M}T_{k,n_1,m}}{M}$.

According to the Central Limit Theorem, when the sample size is sufficiently large, regardless of the distribution of the population, $\overline{T_{k,n_1}}$ follows a normal distribution. Thus,

\begin{equation}
    \begin{aligned}
        \mu &= \lim\limits_{n \to \infty}E_{H_0}(T_{k,n_1,m}) 
        = \lim\limits_{n \to \infty} \frac{n_2}{n-1} = \lambda_2
    \end{aligned}
    \end{equation}
    \begin{equation}
    \begin{aligned}
        \sigma^2 &= \lim\limits_{n \to \infty}Var_{H_0}(T_{k,n_1,m}) \\
        &= \lambda_2^2\frac{\sum\limits_{r=1}^{k}\sum\limits_{s=1}^{k}p_1(r,s)}{k^2}\\
        &+\lambda_1 \lambda_2 \frac{\sum\limits_{r=1}^{k}\sum\limits_{s=1}^{k}p_2(r,s)}{k^2} \\
        % &+\frac{n_2(nn_1-2n+1+n_2-knn_1-kn_1n_2+3kn_1+2kn_2-2k)}{kn_1(n-1)^2(n-2)}
    \end{aligned}
    \end{equation}

\end{proof}

}

\input{main.bbl}

% \bibliographystyle{IEEEtran}
% \bibliography{myref}

\end{document}